\documentclass[12pt]{article}
\pdfoutput=1
\usepackage{epsfig}
\usepackage{amsfonts}
\usepackage{amssymb}
\usepackage{xcolor}
\usepackage{titlesec}
\usepackage{multirow}
\usepackage{bm}
\usepackage{amsmath}
\usepackage{comment}
\usepackage{silence}
\usepackage{mathtools}
\usepackage{cancel}
\usepackage{color}
\usepackage{slashed}
\usepackage{cite}
\usepackage{calligra}
\usepackage{caption}
\usepackage{subcaption}
\usepackage{marginnote}
\usepackage{accents}

\titleformat{\paragraph}
{\normalfont\normalsize\bfseries}{\theparagraph}{1em}{}
\titlespacing*{\paragraph}
{0pt}{3.25ex plus 1ex minus .2ex}{1.5ex plus .2ex}
\begin{document}
\newcommand{\nc}{\newcommand}
\nc{\beq}{\begin{equation}} \nc{\eeq}{\end{equation}}
\nc{\beqa}{\begin{eqnarray}} \nc{\eeqa}{\end{eqnarray}}
\nc{\eps}{{\epsilon}}
\nc{\R}{{\cal R}}
\nc{\A}{{\cal A}}
\nc{\K}{{\cal K}}
\nc{\B}{{\cal B}}
\nc{\C}{{\cal C}}
\nc{\N}{{\cal N}}
\begin{center}

{\bf \Large   UV Divergences, RG Equations and High Energy Behaviour of the Amplitudes in  the Wess-Zumino \\ [0.3cm] Model with Quartic Interaction} \vspace{1.0cm}

{\bf \large L. V. Bork$^{1,2}$ and D. I. Kazakov$^{3,4}$ }\vspace{0.5cm}

{\it
$^1$Institute for Theoretical and Experimental Physics, Moscow, Russia\\
$^2$Center for Fundamental and Applied Research, All-Russian Institute of Automatics, Moscow, Russia \\
$^3$Bogoliubov Laboratory of Theoretical Physics, Joint Institute for Nuclear Research, Dubna, Russia.\\
$^4$Moscow Institute of Physics and Technology, Dolgoprudny, Russia}
\vspace{0.5cm}

\abstract{
 We analyse the UV divergences for the scattering amplitudes in the Wess-Zumino SUSY model with the quartic superpotential. We argue that the only divergent amplitudes are those with four external legs. Within the superfield formalism, we calculate  the corresponding Feynman diagrams and evaluate their leading divergences up to 4 loop order of PT. Then we construct recurrence relations that connect the leading UV divergences in subsequent orders of perturbation theory. These recurrence relations allow us to calculate the leading divergences in a pure algebraic way starting from the one loop contribution. We check that the obtained relations correctly reproduce the lower order diagrams evaluated explicitly. At last, we convert the recurrence relations into the RG equations that have integro-differential form. Solving these equations for a particular sequence of diagrams, we find out the  high energy behaviour of the amplitude. We then argue that the full amplitude has a similar behaviour with the key feature of the existence of a pole in the s-channel corresponding to a state with a mass $\sim 1/g$, where $g$ is the original dimensionfull coupling of the theory. We find out the this state is actually a ghost one similar to the Landau pole in scalar theory.}
\end{center}


\newpage

\tableofcontents{}\vspace{0.5cm}

\section{Introduction}\label{Intro}

Non-renormalizable interactions continue to attract attention partly due to their role in effective theories 
describing potentially new physics at high energy scale and partly due to incessant attempts to construct a consistent theory of quantum gravity. 

In a series of papers~\cite{Bork:2014nma,Bork:2015zaa,Kazakov:2016wrp,Borlakov:2016mwp,Borlakov:2017vra}, \cite{Kazakov:2018zgi,Kazakov:2019wce},  we demonstrated how one can deal with non-renorma\-lizable quantum field theory, trace and sum all types of logarithmic ultraviolet (UV) divergences (leading, subleading, etc.), and write down the generalized renormalization group (RG) equations that enable one to describe the high energy behaviour of the scattering amplitudes in full analogy with the renormalizable case.
As a playground of our analysis we considered the planar maximally supersymmetric Yang-Mills gauge theories (SYM) in 6, 8 and 10 dimensions \cite{Bork:2014nma,Bork:2015zaa,Kazakov:2016wrp,Borlakov:2016mwp,Borlakov:2017vra}  as well as the $\phi^4$ scalar theory\cite{Kazakov:2018zgi,Kazakov:2019wce} in $D$ dimensions. The object of studies was the four point scattering amplitude on mass shell. The choice of such theories as well as observables therein was motivated mostly by the availability of explicit representation of the four-point amplitudes up to some high order of perturbation theory (PT) in terms of master integrals. 
 It is also worth mentioning that gauge theories with extended supersymmetry in $D=4$ possess unique properties among  $D=4$ QFTs such as integrability (see \cite{Beisert:2010jr} and references therein) and explicit all loop expressions for some scattering amplitudes and correlation functions \cite{BDS,Henrietta_Amplitudes}. One can hope that higher dimensional counterparts of these supersymmetric gauge theories will also reveal rich and interesting structures. 

In the present paper we focus our attention on a four-dimensional example, namely, we consider a supersymmetric model of the Wess-Zumino type with quartic interaction that is formally non-renormalizable.
 We analyse the UV divergences for the scattering amplitude in this model within the usual PT in the superfield formulation, reproduce the corresponding Feynman diagrams and evaluate their leading divergences up to 4 loop order of PT. Then we construct recurrence relations that connect the leading UV divergences in subsequent orders of perturbation theory. These recurrence relations allow us to calculate the leading divergences in a pure algebraic way starting from the one loop contribution. We check that the obtained relations correctly reproduce the lower order diagrams evaluated explicitly. At last, we convert the recurrence relations into RG equations that have integro-differential form. Solving these equations for a particular sequence of diagrams, we find out the  high energy behaviour of the amplitude. We then argue that the full amplitude has a similar behaviour with the key feature of the existence of a pole in the s-channel corresponding to a state with a mass $\sim 1/g$, where $g$ is the original dimensionfull coupling of the theory. We find out the this state is actually a ghost one similar to the Landau pole in  scalar theory.

\section{The UV divergences of the four point scattering amplitude in the Wess-Zumino type model}
\label{sWZ}

Consider a supersymmetric model of the Wess-Zumino type with the Lagrangian
\begin{eqnarray}
\label{LWZ}
{\cal L}= \int d^2\theta d^2\bar \theta\  \bar \Phi \Phi +\int d^2\bar \theta \  \frac{g}{4!} \Phi^4
+\int d^2 \theta \  \frac{g}{4!}\bar \Phi^4, \label{lag}
\end{eqnarray}
where $\Phi(x,\theta,\bar{\theta})$ and $\bar \Phi(x,\theta,\bar{\theta})$ are the $D=4$ scalar chiral and antichiral $N=1$ superfields. In what follows, we consider the massless case for simplicity. This allows us to perform analytical calculation of the corresponding diagrams.

The on-shell states which are contained in the chiral superfields $\Phi$ and
$\bar \Phi$ are given by the scalar field $\phi$ and the chiral fermion field $\psi_+$ with positive helicity and the scalar field $\phi^*$ and the chiral fermion fermion field $\bar \psi_-$ with negative helicity,  respectively. 
Written in terms of these  fields the Lagrangian (\ref{lag}) contains interactions of the form
\begin{equation}
 g\ \psi \psi \phi \phi \ \ \ \mbox{and}\ \ \  g^2 \phi^6,
\end{equation}
which obviously belong to the non-renormalizable type with the coupling $g$ of a mass dimension -1. In the massless case, this will be the only scale in the model. 

In what follows, we consider the four-point scattering amplitude on mass shell. To evaluate the amplitudes,
it is useful to use modern unitarity based methods in on-shell momentum superspace (see \cite{Henrietta_Amplitudes,Talesof1001Gluons,Britto:2010xq} and references therein). However, in this particular case, it is easier to use the superspace Feynman rules and calculate the effective action $\Gamma[\Phi,\bar\Phi]$.

Indeed, the structure of all four-point amplitudes $A_4$ in this model to all loop orders can be schematically written as:
\begin{equation}\label{schemAmpl}
A_4=(\mbox{Polarisation factor}) \times (\mbox{Universal scalar function}~C~\mbox{or}~M),
\end{equation}
where we distinguish the  pure chiral ($C$) and mixed ($M$) cases.
The polarisation factor\footnote{In the pure chiral case, the polarisation factor is given by the tree level amplitude.} depends on the helictites of the particles and is irrelevant for our discussion, while the universal scalar functions $C$ and $M$  are identical for all particles and will be the object of our interest. 
They depend on the Mandelstam invariants $s,t,u$, the coupling constant $g$ and the UV regulator. We will often refer to them as amplitudes themselves. The explicit form of the polarisation factor can be found in Appendix \ref{ap2} in terms of the on-shell momentum superspace variables \cite{Henrietta_Amplitudes}.

The functions $C$ and $M$ in turn can be extracted from the strongly connected parts of the correlation functions of the chiral and anti-chiral superfields
\beq
\langle \Phi \Phi \Phi \Phi \rangle, \ \  \langle \bar\Phi \bar \Phi \bar\Phi \bar\Phi \rangle, \ \ \langle \bar\Phi \bar\Phi \Phi \Phi \rangle.
\eeq
after imposing the on-shell constraints on the external superfields 
\beq
\bar D^2 \bar \Phi=0,  \ \  D^2 \Phi=0,
\eeq
where $D$ and $\bar D$ are the corresponding supercovariant derivatives 
\beq
D_\alpha =\frac{\partial}{\partial\theta^\alpha}+i \frac{1}{2}\bar \theta^{\dot{\alpha}}\frac{\partial}{\partial x^{\alpha \dot \alpha}}, 
\ 
\bar D_{\dot{\alpha}} =\frac{\partial}{\partial\bar\theta^{\dot{\alpha}}}-i \frac{1}{2}\theta^{\alpha}\frac{\partial}{\partial x^{\alpha \dot \alpha}}
\eeq
and $D^2=1/4~D^\alpha D_\alpha$ and $\bar D^2 = 1/4~\bar D^{\dot{\alpha}} \bar D_{\dot{\alpha}}$ (see for \cite{Gates:1983nr} conventions)
and  $p_i^2=0$ condition for external momenta.

\begin{figure}[ht]
 \begin{center}
  \epsfxsize=10cm
 \epsffile{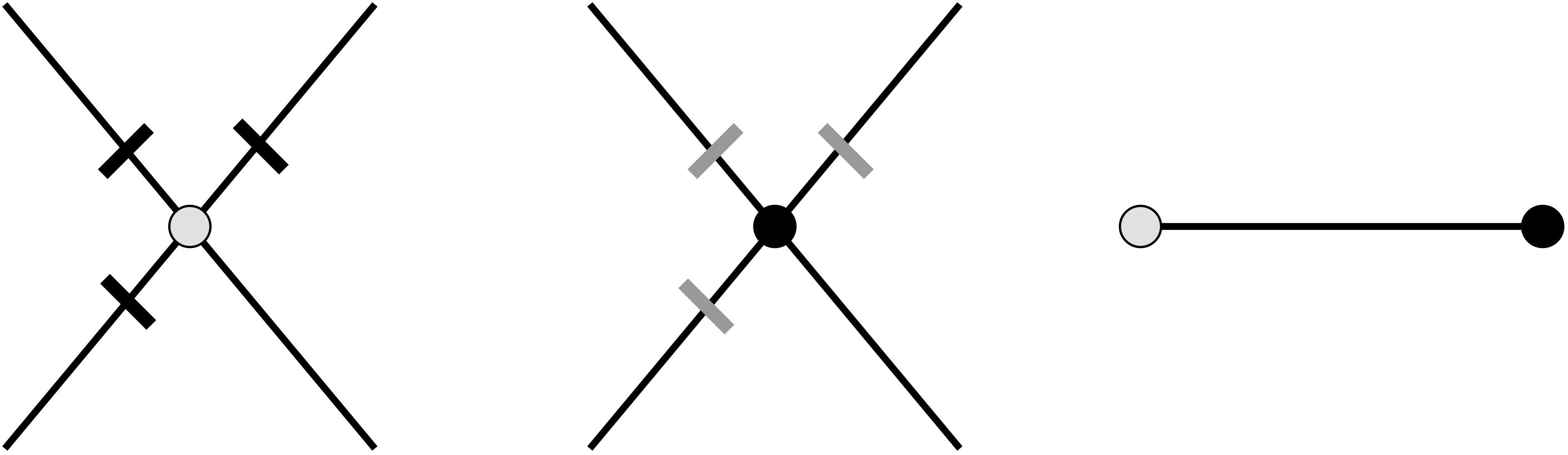}
 \end{center}\vspace{-0.2cm}
 \caption{Feynman rules for the massless WZ model with the quartic superpotential. The thick bar on the line corresponds to the $\bar D^2$ factor, the thin bar corresponds to $D^2$. When the line is external, no $D$ factor on this line is present. The propagator connects only chiral - antichiral vertices and is given by $i\delta^2(\theta )\delta^2(\bar\theta)/p^2$.}\label{figFRules}
 \end{figure}

At each loop order the strongly connected parts of the correlation functions $\langle \Phi_1 \Phi_2 \Phi_3 \Phi_4 \rangle$ and $\langle \Phi_1 \Phi_2 \bar\Phi_3 \bar\Phi_4 \rangle$ can be extracted from the effective action $\Gamma[\Phi,\bar\Phi]$ and are given by the sum of supergraphs which, after the $D$-algebra, can be simplified to expressions that are local in $\theta$ space \cite{Gates:1983nr}. Namely, they can be written as:
\begin{eqnarray}\label{f1}
\langle \Phi_1 \Phi_2 \Phi_3 \Phi_4 \rangle \sim \int d^2\theta ~\prod_{i=1}^4d^4p_i ~\Phi(p_i,\theta)~
C(s,t,u,g),\label{f1}\\
\langle \bar \Phi_1 \bar\Phi_2 \bar\Phi_3 \bar \Phi_4 \rangle \sim \int d^2\bar \theta ~\prod_{i=1}^4d^4p_i ~\bar\Phi(p_i,\theta)~
\bar C(s,t,u,g)\label{f2},
\end{eqnarray}
and
\begin{eqnarray}\label{f3}
\langle \Phi_1 \Phi_2 \bar\Phi_3 \bar\Phi_4 \rangle \sim \int ~d^4\theta~\prod_{i=1}^2d^4p_i\Phi(p_i,\theta)\prod_{i=3}^4d^4p_i\bar \Phi(p_i,\bar \theta)~
MS(s,t,u,g)
\end{eqnarray}
for the chiral and the mixed amplitudes, respectively. Here $MS$ stands for the mixed amplitude $M$ in the 
s-channel. The mixed amplitudes in the t- and u-channels $MT$ and $MU$ are related to the $MS$ one via crossing symmetry, respectively,
\beq
MT(s,t,u,g)=MS(t,u,s,g),~MU(s,t,u,g)=MS(u,s,t,g).
\eeq
Note that due to the condition $p_i^2=0$ each amplitude depends on the Mandelstam variables $s,t$ and $u$ only.
The relevant Feynman rules are presented in fig.\ref{figFRules}. All the contributing supergraphs and the corresponding Feynman integrals  up to the four loop order are given in Appendix A.
It is also convenient to consider the sum of the  $MS$, $MT$ and $MU$ amplitudes 
\beq
M(s,t,u,g)=MS(s,t,u,g)+MT(s,t,u,g)+MU(s,t,u,g).
\eeq
Will refer to it just as the mixed amplitude $M$.

Note also that the Grassmannian integration in (\ref{f1},\ref{f2}) is $d^2\theta$ (or $d^2\bar \theta$) and not $d^2\theta d^2\bar\theta$. This is not an accident but reflects a somewhat unexpected property of the Wess-Zumino model with quartic interaction. The usual non-renormalization theorem in this case is valid, but the chiral part of the effective action is still renormalised. We consider this issue in more detail in the next section. 

Thus, in the model with quartic interaction one has divergent contributions to the four-point Green functions of two types: the chiral (or antichiral) and the mixed ones. They are given by PT series over $g^2$
\beq
C(s,t,u,g)=\frac{g}{4!}\sum_{l=0}~g^{2l}~C^{(l)}(s,t,u), \ \ \ M(s,t,u,g)=\frac 14\sum_{l=1}~g^{2l}~M^{(l)}(s,t,u).
\label{fun}
\eeq
However, due to the Feynman rules in superspace, where the propagator of the massless field is always mixed
$$\langle\Phi \bar\Phi\rangle= i\frac{\delta^2(\theta)\delta^2(\bar\theta)}{p^2}, \ \  \langle\Phi \Phi\rangle=0, \ \ 
\langle\bar \Phi \bar\Phi\rangle=0,$$
one obtains contributions to the chiral vertex only in the even number of loops and to the mixed vertex in the odd ones. We present  individual contributions of the superfield Feynman diagrams up to 4 loops in appendix \ref{ap1}.

It is also convenient to uniformly separate the $s$, $t$ and $u$-channel contributions to the $C$ and $M$ amplitudes which are given by the identical Feynman integrals and differ only by external momenta relabelling. The total contribution is the sum of all three channels
\beqa
C^{(l)}(s,t,u)&=&CS^{(2l)}(s,t,u)+CT^{(2l)}(s,t,u)+CU^{(2l)}(s,t,u), \\
M^{(l)}(s,t,u)&=&MS^{(2l+1)}(s,t,u)+MT^{(2l+1)}(s,t,u)+MU^{(2l+1)}(s,t,u),
\eeqa
where, due to crossing symmetry, the contributions of all channels can be expressed via $CS_{2l}$  and 
$MS_{2l+1}$ by cyclic change of the arguments
\beqa
CT^{(2l)}(s,t,u)&=&CS^{(2l)}(t,u,s),~CU^{(2l)}(s,t,u)=CS^{(2l)}(u,s,t),\\
MT^{(2l+1)}(s,t,u)&=&MS^{(2l+1)}(t,u,s),~MU^{(2l+1)}(s,t,u)=MS^{(2l+1)}(u,s,t),
\eeqa

To regularise the UV divergences within the superfield formalism, we use dimensional regularisation which is very convenient when calculating multiloop Feynman integrals despite the obvious contradiction with supersymmetry. To cure this problem, the modified dimensional regularisation, the so-called dimensional reduction \cite{Capper:1979ns, Siegel:1979wq}, is used which is still not completely adequate \cite{Avdeev:1982xy} but is safe when considering the leading divergences.
Then the logarithmic UV divergences manifest themselves as poles $1/\epsilon^n$ when integrating in $D=4-2\epsilon$ dimensions. All the Grassmannian algebra is performed in 4 dimensions. 

The UV divergent coefficients are always polynomials in $s$,$t$ and $u$ due to the locality of the theory. Since hereafter we focus on the leading UV divergences only, we rewrite the divergent contributions as
\beq
C^{(2l)}(s,t,u)=\frac{C_{2l}(s,t,u)}{\epsilon^{2l}},  \ \ M^{(2l+1)}(s,t,u)=\frac{M_{2l+1}(s,t,u)}{\epsilon^{2l+1}}, \ etc.
\eeq
The functions $CS_{2l}(s,t,u)$ and $MS_{2l+1}(s,t,u)$ are now given by the homogeneous polynomials of $s$, $t$ and $u$  of degree $l$ and $l-1$, respectively. 
Substituting the explicit results for the leading UV poles for the first several orders of PT, we have:
\beqa
C(s,t,u,g)&=&\frac{g}{4!}  \left\{ 1+\frac{g^2}{4}[\frac{s}{\epsilon^2}+\frac{t}{\epsilon^2}+\frac{u}{\epsilon^2}]+\frac{g^4}{32}[\frac{s^2}{\epsilon^4}+\frac{t^2}{\epsilon^4}+\frac{u^2}{\epsilon^4}]+ ...\right\}=\bar C\\
M(s,t,u,g)&=&\frac{1}{4} \left\{\frac{g^2}{2} [\frac 1\epsilon+\frac 1\epsilon+\frac 1\epsilon] \right.\\ &&\left.+g^4\left[
\frac{s}{8\epsilon^3}+\frac{t}{8\epsilon^3}+\frac{u}{8\epsilon^3}+\left(-\frac s2 \frac{1}{3\epsilon^3}-\frac t2 \frac{1}{3\epsilon^3}-\frac u2 \frac{1}{3\epsilon^3}\right)  \right.\right.\nonumber\\&&\left. \left.+
\left(\frac s2 \frac{1}{3\epsilon^3}+\frac t2 \frac{1}{3\epsilon^3}+\frac u2 \frac{1}{3\epsilon^3}\right) \right]+...\right\}\nonumber
\eeqa
Note that the chiral(antichiral) amplitudes contain only the even leading poles while the mixed amplitude contains only the odd ones. A full list of $CS_{2l}$ and $MS_{2l+1}$ up to $l=2$ (4 loops) can be found in appendix \ref{ap1}. 
It is also interesting to note that the leading $1/\epsilon^2$ and $1/\epsilon^3$ poles are actually absent in the $C$ and $M$ amplitudes, respectively, due to the on-shell condition $s+t+u=0$.

\section{Non-renormalisation theorems for arbitrary superpotential}
\label{sNRtheor}

In this section, we discuss in more detail the structure of divergences of the four-point correlation functions of chiral superfields in the Wess-Zumino model with an arbitrary superpotential. We concentrate in particular on radiative corrections to the superpotential which is a novel feature of the WZ model with quartic interaction contrary to the cubic one. The key issue is the so-called non-renormalisation theorem~\cite{Gates:1983nr, West:1990tg} known since the early days of the $N=1$ superspace formalism.

The statement is that any contribution to the $N=1$ superspace effective action for any local theory with chiral and vector superfields at each order of PT is local in the $\theta$ space. This means that in a theory with chiral superfields only it can be written as \cite{West:1990tg}:
\begin{eqnarray}\label{theor1}
\Gamma_n[\Phi,\bar\Phi]=\int \!d^4\theta\prod_{i=1}^nd^4p_iF(\Phi(p_i,\theta),\bar\Phi(p_i,\bar\theta),D^\alpha\Phi(p_i,\theta),\bar D_{\dot{\alpha}}\bar\Phi(p_i), ...)~\mathcal{F}_n(p_1,\ldots,p_n),
\end{eqnarray}
where $d^4\theta=d^2\theta d^2\bar\theta$, $F$ is a local function of $\Phi,\bar \Phi$ and supercovariant derivatives $D^{\alpha},\bar D_{\dot{\alpha}}$, and $\mathcal{F}_n(p_1,\ldots,p_n)$ is the function of only bosonic variables proportional to $\delta^4(\sum_{i=1}^np_i)$ 

This theorem is a direct consequence of the $N=1$ superspace Feynman rules, and the sketch of the proof of this theorem is the following: at each order of PT $\Gamma_n[\Phi,\bar\Phi]$ is given by a finite number of  the $N=1$ superspace Feynman diagrams. Each $N=1$ superspace Feynman diagram is constructed from the propagators which are proportional to the full fermionic delta function  $\delta^4(\theta_i-\theta_{i+1})$ (here $\delta^4(\theta)=\delta^2(\theta)\delta^2(\bar\theta)$) and the vertices which contain supercovariant derivatives $D^2$ or $\bar D^2$ acting  on adjacent propagators, and integration over the full $N=1$ superspace 
$d^4\theta_i=d^2\theta_id^2\bar\theta_i$.
Then, using integration by parts, the covariant derivatives $D^2$ and $\bar D^2$ from the vertices can be rearranged into the combinations such as $\delta^4(\ldots)[D^2\bar D^2\delta^4(\ldots)]$ which can be simplified according to the following identity: 
\begin{equation}
\delta^4(\theta_i-\theta_{i+1})[D^2\bar D^2\delta^4(\theta_i-\theta_{i+1})]=\delta^4(\theta_i-\theta_{i+1}),
\end{equation}
and similar ones.
This in turn allows one to remove the integration with respect to $d^4\theta_{i+1}$ and replace all $\theta_{i+1}$ with $\theta_i$. Carefully analysing the superspace Feynman diagram,  one can show that all but one integration with respect to $d^4\theta$ can be removed this way so that the remaining expression will be local in $\theta$ and can be represented as a single integral over the full $N=1$ superspace. The result for the sum of all $N=1$ superspace Feynman diagrams can then be written as (\ref{theor1}). 

An immediate  consequence of this theorem is that in the theory with the Lagrangian 
\begin{eqnarray}
\label{LWZ_1}
{\cal L}= \int d^4\theta\  \bar \Phi \Phi +\int d^2\bar \theta \  {\cal W}(\Phi)
+\int d^2 \theta \  {\cal W}(\bar \Phi), 
\end{eqnarray}
with
\begin{equation}\label{W_qube}
{\cal W}(\Phi)=\frac{1}{2!}m\Phi^2+\frac{1}{3!}g\Phi^3
\end{equation}
the counter terms proportional to $\int d^2\theta ~\mathcal{W}(\Phi)$ are forbidden since it is the integral over the chiral superspace while the radiative corrections are the integrals over the full superspace. Hence the  superpotential  does not receive "divergent" radiative corrections and is not renormalized. This statement is known as the non-renormalization theorem.

There is, however,  a possible loophole in this reasoning if one talks about the chiral part of the effective action. It is based on the possibility to rewrite the
integration over the full superspace $d^4\theta$  as an integral over the chiral subspace 
as $d^4\theta=d^2\theta \bar D^2$~\cite{West:1990tg} so that the integral for the chiral part of the effective action takes the  form:
\begin{eqnarray}\label{d4_to_d2}
\Gamma_n[\Phi]=\int d^2\theta \bar D^2 \prod_{i=1}^nd^4p_iF(\Phi(p_i,\theta),D^\alpha\Phi(p_i,\theta), ...)~\mathcal{F}_n(p_1,\ldots,p_n)
\end{eqnarray}
and would be equal to zero unless the function F contains covariant derivatives. 

This possibility is ruled out in any renormalizable local theory since all the counter terms must be local and cannot contain powers of momenta in the denominator which is needed to compensate the covariant derivative 
that has a mass dimension of 1/2. And there is no any other scale in the theory except for the mass, but it cannot stay in the denominator in the UV counter terms as well. 

However, in the non-renormalizable case, the situation is different. Now the coupling constant has a negative mass dimension and may act as a needed scale. Then, in the presence of covariant derivatives, the action of  $\bar D^2$ from the integration measure on $D^2 \Phi(p_1)\Phi(p_2)...$ gives
\beq\label{d_to_p}
\bar D^2 D^2 ( \Phi(p_1)\Phi(p_2)...)=-(p_1+p_2+...)^2(\Phi(p_1)\Phi(p_2)...)
\eeq
and one is left with the chiral integral in eq.(\ref{d4_to_d2}) which is not vanishing.  Thus, the radiative corrections to the chiral part of the effective action are quite possible, though they depend on momenta, which in the coordinate space correspond to spatial derivatives. From this point of view, the effective potential remains untouched, though the chiral part of the effective action obtains divergent corrections.  Note also that according to (\ref{d4_to_d2}) the \emph{finite} corrections to the superpotential are allowed in both  renormalizable and non-renormalizable case since there are no locality restrictions on the finite contributions \cite{West:1990rm,Buchbinder:1994xq}. 
Indeed, one can see that using the relation $d^4\theta=d^2\theta \bar D^2$ and (\ref{d4_to_d2}), one can  transform the integral over the  full superspace with the non-local integrand into the integral over the chiral subspace:
\begin{equation}
\int d^4\theta~ f(\Phi)\frac{D^2}{-Q^2}g(\Phi)=\int d^2\theta~ f(\Phi)g(\Phi),
\end{equation}
where $f$ and $g$ are arbitrary functions of the chiral superfields $\Phi$, and $Q$ is the total momentum of the function $g(\Phi)$.  As was shown in \cite{Buchbinder:1994xq}, such expressions indeed appear in explicit calculations in the WZ model with cubic interaction.

We demonstrate below how the generation of the radiative corrections to the divergent chiral part of the effective action happens on example of the WZ model with the quartic superpotential (see also discussion in \cite{Petrov:2021ddc}).
Let us consider the first non-vanishing loop correction to the strongly
connected part of  the chiral correlation function.
\begin{figure}[ht]
 \begin{center}
  \epsfxsize=7cm
 \epsffile{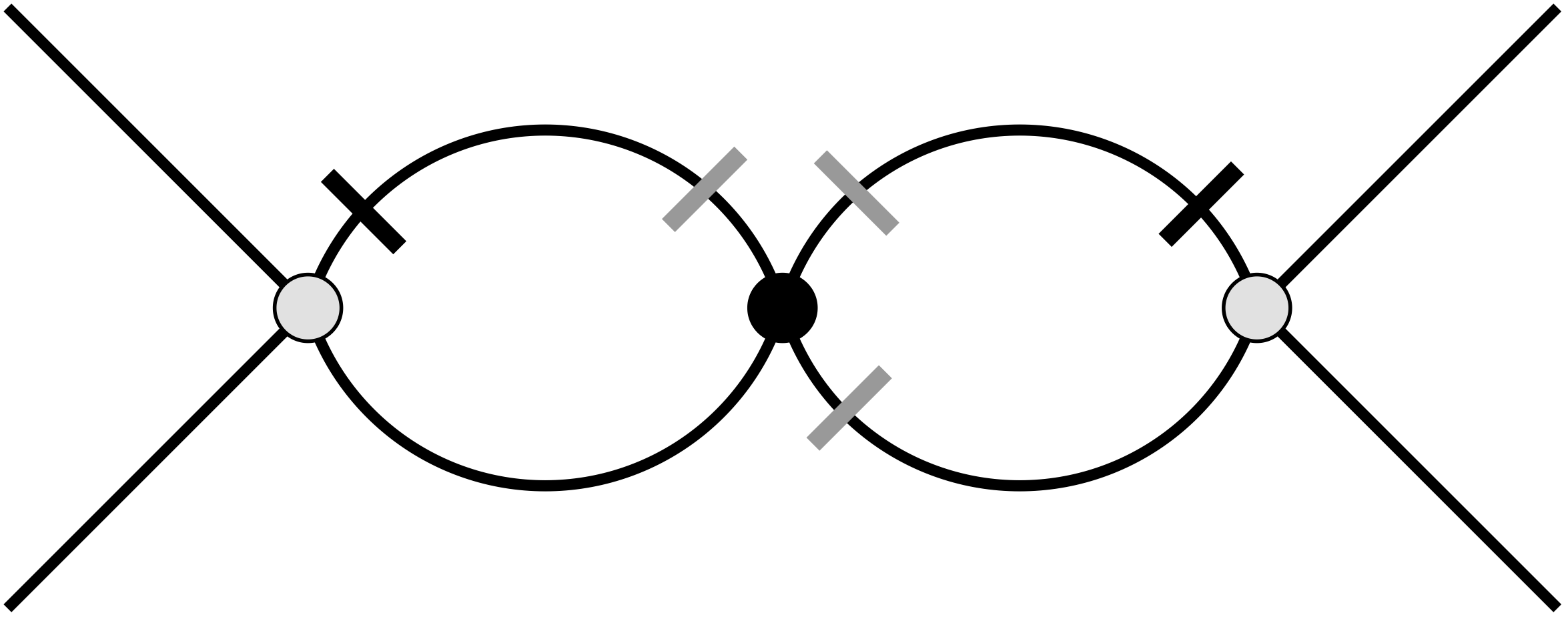}
 \end{center}\vspace{-0.2cm}
 \caption{First none vanishing loop correction to $\langle \Phi \Phi \Phi \Phi \rangle$. Only the $s$-channel diagram is shown.}\label{figNRtheorem}
 \end{figure}
The contribution to the effective action of the two-loop $s$-channel supergraph presented in Fig.\ref{figNRtheorem} is given by
($\delta^4(\theta_i-\theta_j)\equiv \delta_{ij}$, $p_i+p_j\equiv p_{ij}$):
\beq
\Gamma_4^{(1)}=g^3\int \prod_{i=1}^4 d^4\theta_i d^4p_i\Phi_i(\theta_i)
~\delta^4(\sum_{i=1}^4p_i)
\int \frac{d^Dl_1~\delta_{12}[\bar D^2 D^2 \delta_{12}]}{l_1^2(p_{12}-l_1)^2}\int \frac{d^Dl_2~[D^2\delta_{23}][ D^2 \bar D^2 \delta_{23}]}{l_2^2(p_{12}-l_2)^2}.
\eeq
Using the integration by parts, this expression can be reduced to
\begin{eqnarray}\label{ef1}
\Gamma_4^{(1)}=g^3\int \prod_{i=1}^4d^4p_i~d^4\theta~ \Phi_1(\theta)\Phi_2(\theta)D^2[\Phi_3(\theta)\Phi_4(\theta)]\delta^4(\sum_{i=1}^4p_i)\left(\int \frac{d^Dl}{l^2(p_{12}-l)^2}\right)^2.
\end{eqnarray}
Now we can use $d^4\theta=d^2\theta\bar D^2$ to rewrite this expression as
\begin{eqnarray}
\Gamma_4^{(1)}=g^3\int \prod_{i=1}^4d^4p_i~d^2\theta~ \Phi_1(\theta)\Phi_2(\theta)\bar D^2D^2[\Phi_3(\theta)\Phi_4(\theta)]\delta^4(\sum_{i=1}^4p_i)\left(\int \frac{d^Dl}{l^2(p_{12}-l)^2}\right)^2,
\end{eqnarray}
and using the relation (\ref{d_to_p}), which in our case reduces to
\begin{eqnarray}
\bar D^2D^2[\Phi_i(\theta)\Phi_j(\theta)]=-p_{ij}^2\Phi_i(\theta)\Phi_j(\theta),
\end{eqnarray}
we finally obtain:
\begin{eqnarray}
\Gamma_4^{(1)}=-g^3\int d^2\theta~ \prod_{i=1}^4d^4p_i\Phi_i(\theta)~\delta^4(\sum_{i=1}^4p_i)~p_{34}^2\left(\int \frac{d^Dl}{l^2(p_{12}-l)^2}\right)^2.
\end{eqnarray}
This expression is local (after the subtraction of the lowest order counterterm) and contributes to the renormalization of the superpotential, although with additional powers of momenta. The reason why it became possible is the dimension of the coupling $g$ which is negative: $[g]=-1$ in mass units.  Therefore, $g$ provides an  additional scale that compensates  for super-covariant derivatives.  
Thus,  in the theory with the superpotential $\mathcal{W}(\Phi)\sim \Phi^4$ the chiral counter terms become possible. This is also true for  an arbitrary superpotential $\mathcal{W}(\Phi)$  with higher powers of the fields.

\section{R-symmetry and the structure of the counterterms}

When constructing counterterms,  one should have in mind that Lagrangian (\ref{lag}) possesses the so-called R-symmetry\footnote{See the discussion after $(4.1.15)$ in \cite{Gates:1983nr}.}. Since the regularization does not violate R-symmetry, the counterterms also have to be R-invariant.

The R-symmetry is the global $U(1)$ symmetry of the action under the phase rotation
\beqa && \Phi \to e^{i\beta}\Phi,\  \theta\to e^{2i\beta} \theta,   \  d\theta \to  e^{-2i\beta} d\theta,  \  D_\alpha\to  e^{-2i\beta}D_\alpha, \nonumber\\
&&\bar \Phi \to e^{-i\beta}\bar \Phi, \  \bar \theta\to e^{-2i\beta} \bar\theta,  \   d\bar\theta \to  e^{2i\beta} d\bar\theta,  \  \bar D_{\dot\alpha}\to   e^{2i\beta}\bar D_{\dot\alpha}.
\eeqa

\begin{figure}[ht]
\begin{center} 
\includegraphics[width=0.9\textwidth]{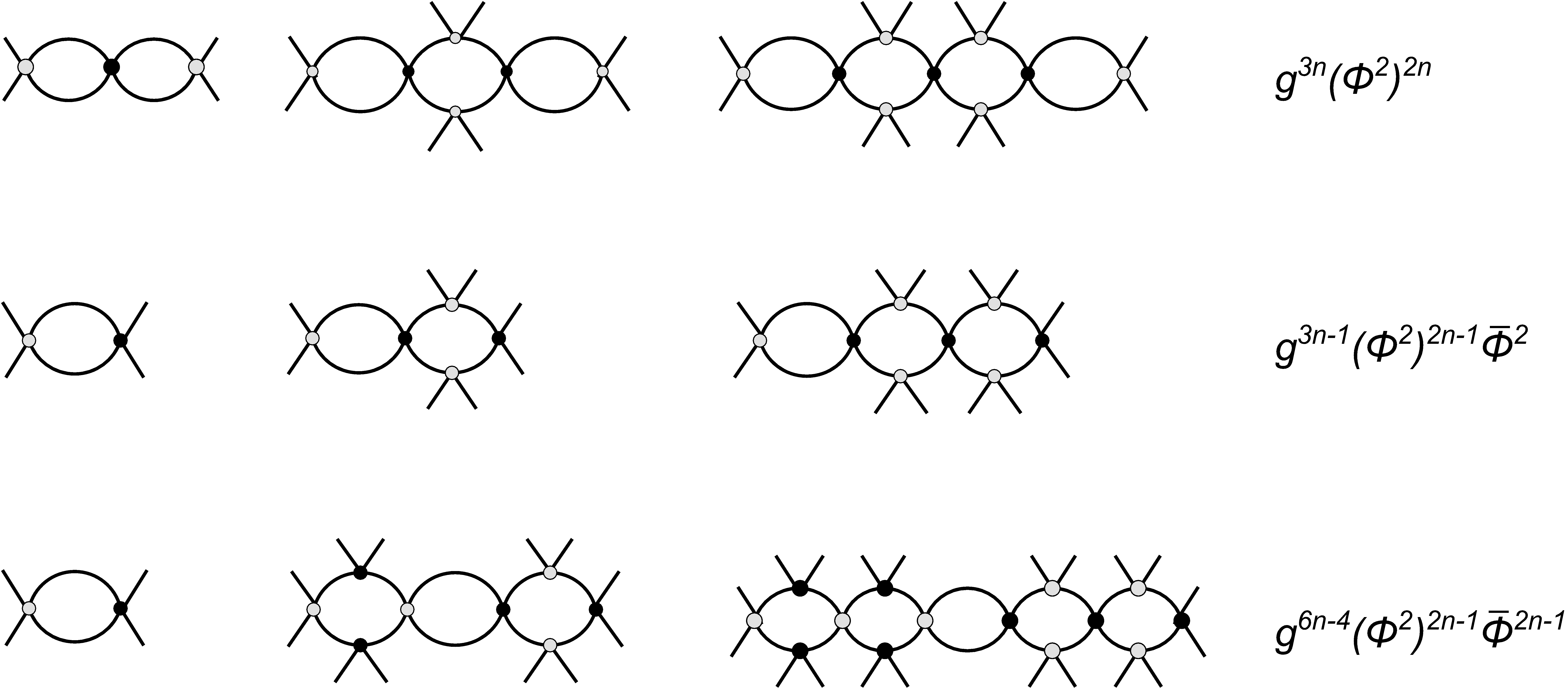}
\caption{Expamples of superfield diagrams with various external legs \label{R-sym}}
\end{center}
\end{figure}

Consider now possible conterterms which appear after the calculation of  Feynman diagrams in the superfield formalism. We distinguish three types of diagrams depending on  the type of external legs:
$$ (\Phi^2)^{2n}, \ (\Phi^2)^{2n-1}\bar\Phi^2, \  (\Phi^2)^{2n-1}(\bar\Phi^2)^{2n-1}., \ \ n=1,2,...$$
There also exist also the conjugated diagrams and intermediate ones, but the chosen set demonstrates all the main features of possible counterterms. Examples of the corresponding  superfield Feynman diagrams are shown in Fig.\ref{R-sym}.

We now have to check the validity of R-symmetry. The obtained expressions have to be integrated over the full superspace $\int d^2 \bar\theta d^2\theta$. The measure itself is R-invariant. Hence, one has to get R-invariant integrands for the sequence of diagrams of Fig.\ref{R-sym}.

We start with the last line. Since in this case one has an equal number of $\Phi$ and $\bar\Phi$, the phases are canceled and R-symmetry is manifest.  For the first and second line this compensation is no more valid, and one has to add a proper number of covariant derivatives to compensate the phase.  Having in mind the Feynman rules where only the squares of covariant derivatives are present, one can expect to have the following expressions:
\beq (D^2\Phi^2)^{n}(\Phi^2)^{n}, \ \  (D^2\Phi^2)^{n}(\Phi^2)^{n-1} \bar D^2\bar\Phi^2,\eeq
which are already R-invariant. (Strictly speaking, the square of the covariant derivative might be splitted into 
two $D_\alpha$, but this does not change the arguments.)

Now let us calculate the dimensions of the integrands. To get a dimensionless effective action, having in mind that the measure of integration has dimension $[d^2\bar\theta d^2\theta]=2$, one should have dimension of the integrand equal to 2. Remind the mass dimensions of the fields and $\theta's$
\beq
[\Phi]=[\bar\Phi]=1, \ [g]=-1,\ [\theta]=[\bar\theta]=-1/2, \ [d\theta= \frac{d}{d\theta}]=1/2, \  [d\bar\theta= \frac{d}{d\bar\theta}]=1/2.
\eeq
We also need the dimensions of covariant the derivatives $D_\alpha$ and $\bar D_{\dot\alpha}$. They are 
$$[D_\alpha]=[\bar D_{\dot\alpha}]=1/2.$$
For a given set of diagrams one has the following integrands, respectively,
$$g^{3n}(D^2\Phi^2)^{n}(\Phi^2)^{n} I_n, \ \
g^{3n-1}(D^2\Phi^2)^{n}(\Phi^2)^{n-1} \bar D^2\bar\Phi^2 I_n, \ \
g^{6n-4}(\Phi^2)^{2n-1}(\bar\Phi^2)^{2n-1}I_n,$$
where $I_n$ is the corresponding ordinary Feynman integral.
Calculating the dimensions, one has
$$ -3n+3n+2n+[I_n]=2 \Longrightarrow [I_n]=2-2n,$$
$$-3n+1+3n+2n-2+3+[I_n]=2 \Longrightarrow [I_n]=2-2n,$$
$$-6n+4+4n-2+4n-2+[I_n]=2 \Longrightarrow [I_n]=2-2n.$$

Thus, we can see that, except for the case of n=1, the Feynman integral $I_n$ always  has a negative dimension, which means that it is globally UV convergent. The only divergent structures are those with four external legs
$$ \Phi^4, \ \ \bar\Phi^4, \ \  \mbox{and} \ \ \Phi^2\bar\Phi^2 .$$
With increasing order of PT, they will acquire  additional powers of the coupling $g^2$ compensated by  powers of the Mandelstam  variables $s,t$ or $u$.
All the other amplitudes with multiple external legs are globally convergent and can only have divergent subgraphs with four external legs.

Therefore, R-symmetry essentially restricts the structure of possible counterterms, and one is left with the chiral, anti-chiral and mixed amplitudes, which are the subject of the present  paper.

\section{Recurrence relations for the leading  divergences}
\label{sRecRel}
Any local  QFT has the property that in higher orders of PT after subtraction of divergent subgraphs,
i.e. after performing the incomplete ${\cal R}$-operation, the so-called ${\cal R}^\prime$-operation, the remaining UV divergences are local functions in the coordinate space or at maximum are polynomials of external momenta in momentum space \cite{Bogolyubov:1980nc11,Zavyalov:1990kv,Vasilev:2004yr}. This follows from the rigorous proof of the Bogoliubov-Parasiuk-Hepp-Zimmermann $\R$-operation~\cite{Bogoliubov:1957gp,Hepp:1966eg,Zimmermann:1969jj} and is equally valid in non-renormalizable theories as well. 

\begin{figure}[ht]
 \begin{center}
  \epsfxsize=12cm
 \epsffile{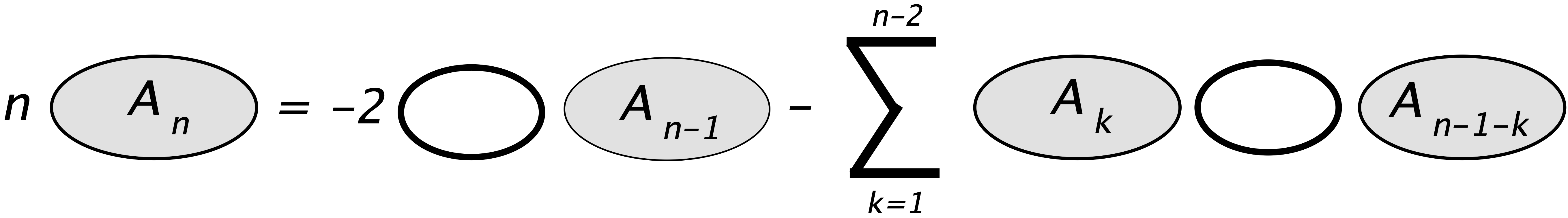}
 \end{center}\vspace{-0.2cm}
 \caption{Recurrence relation for the leading divergences for the four-point amplitude. The white circle corresponds to the  one-loop live diagram remained after the ${\cal R}^\prime$-operation.}\label{recgraf}
 \end{figure}

This property allows one to construct the so-called recurrence relations which relate the divergent contributions in  subsequent orders of perturbation theory.  They allow one to get divergences in all orders of PT starting from  the lower order ones.  In renormalizable theories these relations are known as pole equations (within dimensional regularization) and are governed by the renormalization group \cite{tHooft:1973mfk}. The same is true, though technically is more complicated, in any local theory \cite{Kazakov:1987jp,Bork:2015zaa,Kazakov:2018zgi,Kazakov:2019wce,Buchler:2003vw,Koschinski:2010mr,Polyakov:2018rdp}, renormalizable or not. 
Symbolically, these relations can be written as  shown in Fig.\ref{recgraf} (in the s-channel), where $A_n$ denotes the coefficient of the leading pole in  $n$ loops.

The peculiarity of the procedure is that  in the non-renormalizable case
the pole terms depend on kinematics and are polynomials over momenta. This means that when substituting the counter terms into the recurrence relation, one has to integrate these kinematic factors over the remaining loop \cite{Bork:2015zaa,Kazakov:2018zgi}. This integration can be done in general form by introducing the Feynman parameters. 

Then one arrives at the following recurrence relations for the chiral and mixed amplitudes in the s-channel:
\beqa
&&2n CS_{2n}=\frac 12\left[ 2s \int^1_0 dx MS_{2n -1}(s, t', u')\vert_{t'=-x s, u'=-(1-x)s}\right. \label{r1}\\
&+& \left.2s\int^1_0 dx 
 \sum^{n-1}_{k=1}\sum_{p=0}^{k-1}\sum_{l=0}^{p}\frac{1}{p!p!}  s^p[x(1-x)]^p t^l u^{p-l}
 \frac{d^p}{dt'^l du'^{p-l}} MS_{2k-1}(s, t', u')\times  \right. \nonumber \\ 
&\times &\left.  \frac{d^p}{dt'^l du'^{p-l}} 
\left(CS_{2n-2k}(s,t',u')+CT_{2n-2k}(s,t',u')+CU_{2n-2k}(s,t',u')\right)\vert_{t'=-x s, u'=-(1-x)s}\right]\nonumber \\
&& (2n+1)MS_{2n+1}=\frac 12\left[  
 \int^1_0 dx \sum_{k=0}^{n}\sum_{p=0}^{k-1}\sum_{l=0}^{p}\frac{1}{p! p!}s^p[x(1-x)]^p t^l u^{p-l} \times  \right. \nonumber \\ &&\left. \times \left(\frac{d^p}{dt'^l du'^{p-l}}\bar CS_{2k}\frac{d^p}{dt'^l du'^{p-l}}CS_{2n-2k}+\frac{d^p}{dt'^l du'^{p-l}}\bar CT_{2k}\frac{d^p}{dt'^l du'^{p-l}}CT_{2n-2k} \right. \right.\nonumber \\ && \left. \left.+
 \frac{d^p}{dt'^l du'^{p-l}}\bar CU_{2k}\frac{d^p}{dt'^l du'^{p-l}}CU_{2n-2k}\right)\vert_{t'=-x s, u'=-(1-x)s}
\right.\nonumber\\
&& +\left. s \int^1_0 dx \sum_{k=1}^{n}\sum_{p=0}^{k}\sum_{l=0}^{p}\frac{1}{p! p!}s^p[x(1-x)]^p t^l u^{p-l} \times  \right. \nonumber \\ &&\left. \times \left(\frac{d^p}{dt'^l du'^{p-l}}MS_{2k-1}\frac{d^p}{dt'^l du'^{p-l}}MS_{2n-2k+1}+\frac{d^p}{dt'^l du'^{p-l}}MT_{2k-1}\frac{d^p}{dt'^l du'^{p-l}}MT_{2n-2k+1} \right. \right.\nonumber \\ && \left. \left.+
 \frac{d^p}{dt'^l du'^{p-l}}MU_{2k-1}\frac{d^p}{dt'^l du'^{p-l}}MU_{2n-2k+1}\right)\vert_{t'=-x s, u'=-(1-x)s}\right]\label{r2}
\eeqa
The first linear term of eq.(\ref{r1},\ref{r2})  corresponds to the first diagram in Fig.\ref{recgraf} and the  second nonlinear term is due to the second diagram  with the  live loop in the middle. Integration over $x$ is just the integration over the Feynman parameter in the remaining loop. Multiple sums appear due to the $g_{\mu\nu}$ factors arising when integrating multiple momenta in the numerator of the diagrams in the non-linear term.

Using the recurrence relations (\ref{r1},\ref{r2}), one can evaluate the leading poles in a pure algebraic way starting from the one loop diagram $MS_1$. One gets:
\beqa
&&CS_0=1,\\
&&MS_1=\frac 12,\\
&&2CS_2=\frac 12\left[2s\int_0^1 dx MS_1\right]=\frac 12\left[2s\int_0^1 dx \frac 12\right]=\frac s2 \Longrightarrow CS_2=\frac s4,\\
&&3MS_3=\frac 12\left[\int_0^1 dx \left(CS_2+\bar CS_2+CT_2+\bar CT_2+CU_2+\bar CU_2\right) \right.\nonumber \\
&&\left. +s\int_0^1dx \left(MS_1 MS_1+MT_1MT_1+MU_1MU_1)\right)\right] \nonumber\\
&&=\frac 12\left[\int_0^1 dx \left(\frac s4+ \frac s4+\frac{t'}{4}+\frac{t'}{4}+\frac{u'}{4}+\frac{u'}{4}\right) 
+s\int_0^1dx \left(\frac 12 \frac 12 +\frac 12 \frac 12 +\frac 12 \frac 12 )\right)\right] \nonumber\\
&& =\frac 12\left[\frac{s-s/2-s/2}{2}+3\frac s4\right]=3\frac s8 \Longrightarrow MS_3=\frac s8, \\
&&4CS_4=\frac12\left[2s\int_0^1dx MS_3+2s\int_0^1dx MS_1 (CS_2+CT_2+CU_2)\right] \nonumber \\
&&=\frac12\left[2s\int_0^1dx \frac s8+2s\int_0^1dx \frac 12 (\frac s4+\frac{t'}{4}+\frac{u'}{4})\right] =\frac{s^2}{8}\Longrightarrow CS_4=\frac{s^2}{32},
\eeqa
that coincides with the result of direct calculation of the corresponding diagrams from Table \ref{tab1} and \ref{tab2}. Continuing this procedure we get: 
\beq
CS_0=1, CS_2=\frac s4, CS_4=\frac 12\left(\frac s4\right)^2, CS_6=\frac 59\left(\frac s4\right)^3, CS_8= \frac{61}{126}\left(\frac s4\right)^4, CS_{10}= \frac{718}{1575} \left(\frac s4\right)^5... \label{CS}
\eeq
and
\beqa
MS_1&=&\frac 12, \ MS_3=\frac 12 \frac s4, \ MS_5= \frac{5}{12}\left(\frac s4\right)^2, \ MS_7= \frac{26}{63}\left(\frac s4\right)^3,\label{MS} \\
MS_9&=&(\frac s4)^4\left(\frac{14281}{45360}\!+\!\frac{t}{1080 s}\!+\!\frac{t^2}{1080 s^2}\right),
MS_{11}=(\frac s4)^5\!\left(\frac{773741}{2494800}\!+\!\frac{t}{2376 s}\!+\!\frac{t^2}{2376 s^2}\right)...\nonumber
\eeqa

\section{RG equations and high energy behaviour}
\label{sRGeq}

 The recurrence relations (\ref{r1},\ref{r2})  can be converted into the differential equations which are nothing more than the RG equations in the non-renormalizable case. However, here they are more complicated and inherit the integrals over the Feynman parameters from the recurrence relations. To proceed, we rewrite the  functions $CS$ and $MS$ (\ref{fun}) as
 \beq
 CS=\frac{g}{4!}\sum_{n=0}^\infty CS_{2n} z^{2n}, \ \ \ \  MS=\frac{g}{4}\sum_{n=0}^\infty MS_{2n+1} z^{2n+1}, \ \ \ \   z\equiv \frac g\epsilon   \label{sumz}
 \eeq
 and similar ones in the other channels. 
Then, multiplying the recurrence relation (\ref{r1}) by $z^{2n-1}$ and the recurrence relation (\ref{r2}) by $z^{2n}$ and taking the sum over $n$ from $0$ to $\infty$,  we get the following differential equations
(for the sake of simplicity, in what follows we omit the normalisation factors $g/4!$ and $g/4$ from eq.(\ref{sumz}))
\beqa
\frac{dCS}{dz}&=&s MS \otimes \left(CS+CT+CU\right), \label{eq1} \\
\frac{dMS}{dz}&=&\frac 12[s  (MS \otimes MS+ MT \otimes MT+ MU \otimes MU) \nonumber \\
&&\ \ \ \ \ \ \ +\bar CS\otimes CS +\bar CT\otimes CT+\bar CU\otimes CU], \label{eq2}
\eeqa
with the boundary conditions $CS(0)=1, MS(0)=0$.
Here the symbol $\otimes$ should be understood as
\beqa
&&A(s,t,u) \otimes B(s,t,u) = \int_0^1 dx \sum_{p=0}^{\infty}\sum_{l=0}^{p}\frac{1}{p! p!}\times \\ &&\times \frac{d^p}{dt'^l du'^{p-l}}\ A(s,t',u')\frac{d^p}{ dt'^l du'^{p-l}}\ B(s,t',u')\vert_{\scriptsize  \begin{array}{l}t'=-x s,\\  u'=-(1-x)s\end{array}}\!\!\!\!
s^p[x(1-x)]^p t^l u^{p-l} .\nonumber
\eeqa
This expression with the help of the identity 
\beqa
&&\sum_{p=0}^\infty \frac{(z y)^p k!}{p!(p+k+1)!} \left(\frac{d^p}{dx^p}f(x)\right)^2\\=
&&\frac{1}{2\pi}\int_{-\pi}^{\pi} d\tau \int_0^1 d\xi(1-\xi)^k f(x+\exp(i\tau)\xi~y) f(x+\exp(-i\tau)~z).\nonumber 
\eeqa
can also be rewritten in integral form \cite{Borlakov:2016mwp} which might be useful for numerical integration
\beqa
&&A(s,t,u) \otimes B(s,t,u) =\frac{1}{(2\pi)^2}\int_0^1 dx \int_{-\pi}^{\pi}d\tau\int_{-\pi}^{\pi}d\sigma  \\
&& A(s,-s x\!+\!e^{i\tau}t x,-s(1-x)\!+\!e^{i\sigma}u x) B(s,-s x\!+\!e^{i\tau}s(1- x),-s(1-x)\!+\!e^{i\sigma}s (1-x)).\nonumber
\eeqa

The obtained RG equations allow one to analyse the high energy behaviour of the amplitude though their analytical solution is problematic and one is bounded to rely on numerical approach. It is possible, however, to get a solution for selected series of diagrams. For example, generalising the Lagrangian (\ref{lag}) to the vector $\Phi^a$ or the matrix $\Phi^{ab}$ field, one has the interaction of the form ($a,b=1,...,N$)
$$
 \int d^2\theta \ \frac{g}{4N}(\Phi^a \Phi^a)^2
$$
in the first case and
$$  
\int d^2 \theta \  \frac{g}{4N} (Tr \Phi \Phi)^2 \ \ \mbox{or} \ \ \int d^2 \theta \  \frac{g}{4!N} (Tr \Phi \Phi\Phi\Phi)
$$
in the second case.

For the vector field and the first option of the matrix field the combinatorics is the same and one can consider the $1/N$ expansion. The leading contribution when $N\to\infty$ comes from the sequence of bubbles like in
the ordinary scalar $\phi^4$ theory \cite{Vasilev:2004yr}. On the contrary, for the second option of the matrix field the leading $1/N$
contribution includes all planar diagrams similar to QCD and is much more complicated.

Consider the vector case.
For the sequence of bubbles  $>\!\!\!\bigcirc\!\! \bigcirc \!\!\bigcirc ... \bigcirc\! \!\!<$ in the s-channel one is left with the dependence on $s$ only. Thus, the integrals  as well as the derivatives in eqs.(\ref{eq1},\ref{eq2}) disappear and  RG equations are reduced to familiar differential  form:
\beqa\frac{d CS}{dz}&=& s MS \cdot CS, \label{eq3} \\
\frac{d MS}{dz}&=&\frac 12[ s MS^2+ CS^2], \ \ \ \bar CS=CS \label{eq4}
\eeqa
with the boundary conditions $CS(0)=1, MS(0)=0$.

Solution to eqs.(\ref{eq3},\ref{eq4}) is
\beq
CS(z)=\frac{1}{1-s z^2/4},  \ \ \  MS(z)=\frac{z/2}{1-s z^2/4}, \label{sol}
\eeq
i.e. we get the geometrical progression as expected.  And the same is true for other channels with the replacement $s\to t$ and $s\to u$, respectively. 

\begin{figure}[ht]
  \begin{center}
   \epsfxsize=7.5cm
  \epsffile{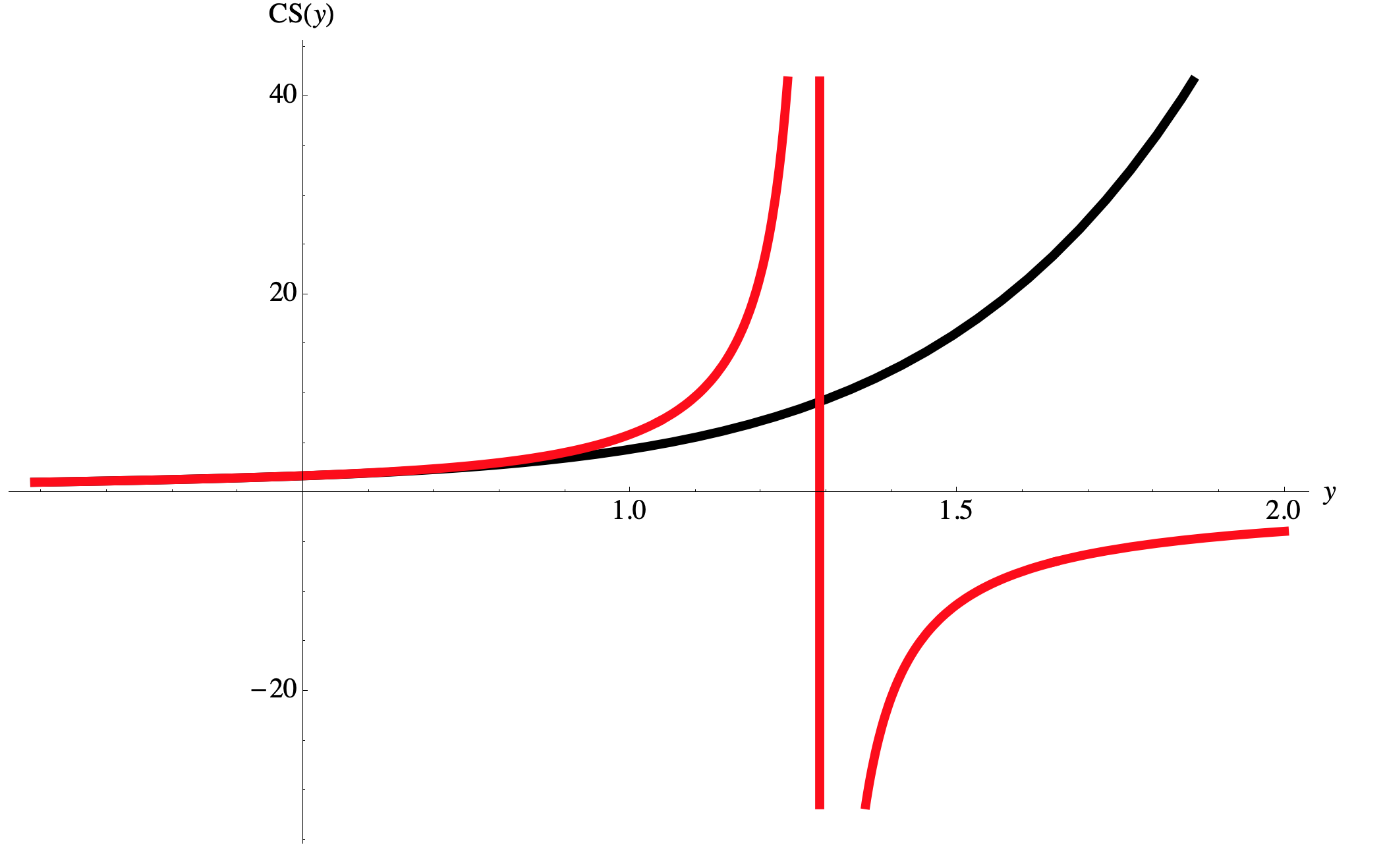}
   \epsfxsize=7.5cm
  \epsffile{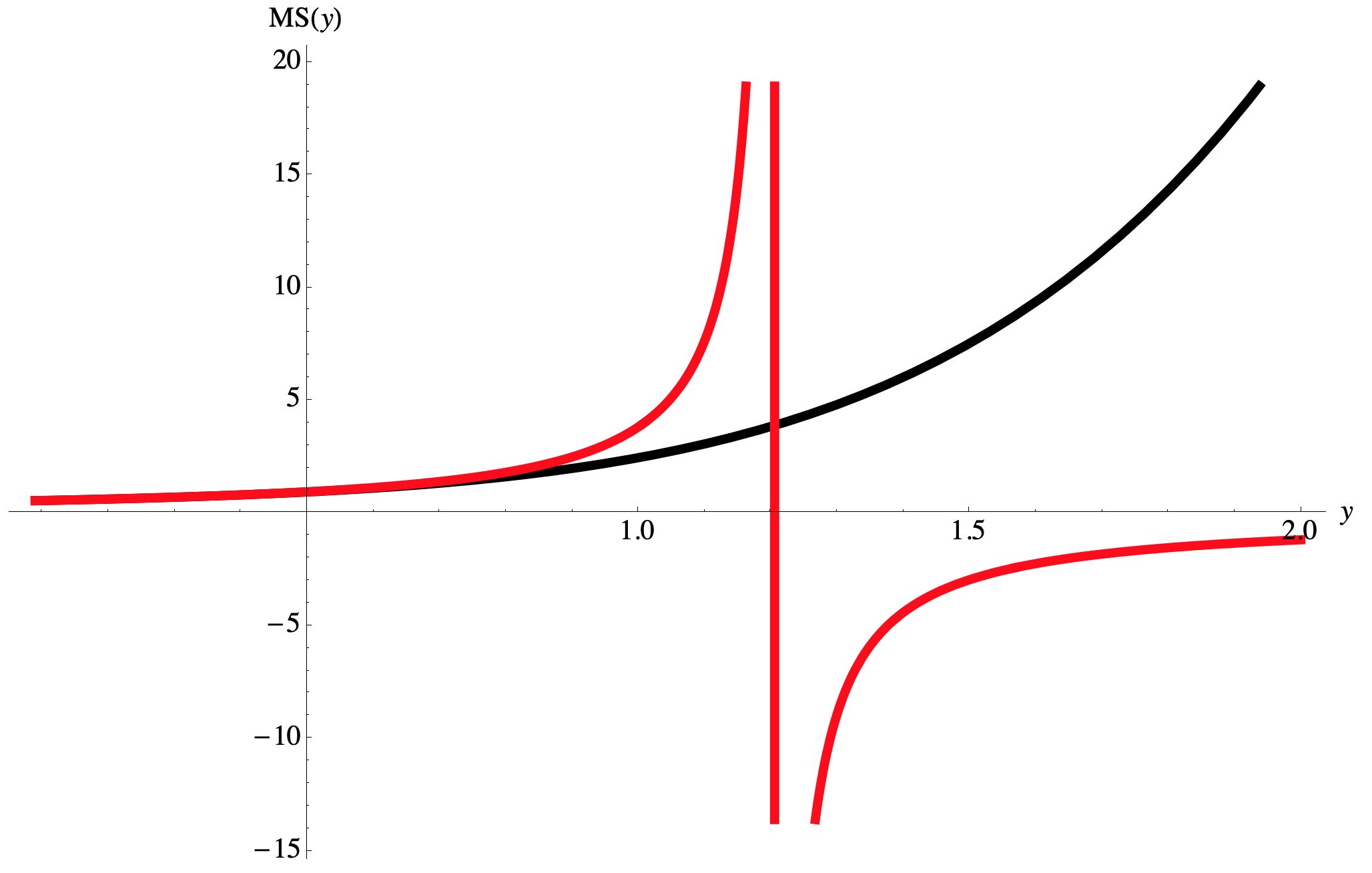}
  \end{center}\vspace{-0.2cm}
  \caption{Plots of the first 6 terms of  expansion for $CS(y)$ (left) and $MS(y)$(right)  shown in black and its  $[9/9]$ Pad\'e approximation (red).}\label{PadeS}
  \end{figure}

\begin{figure}[ht]
  \begin{center}
   \epsfxsize=7.0cm
  \epsffile{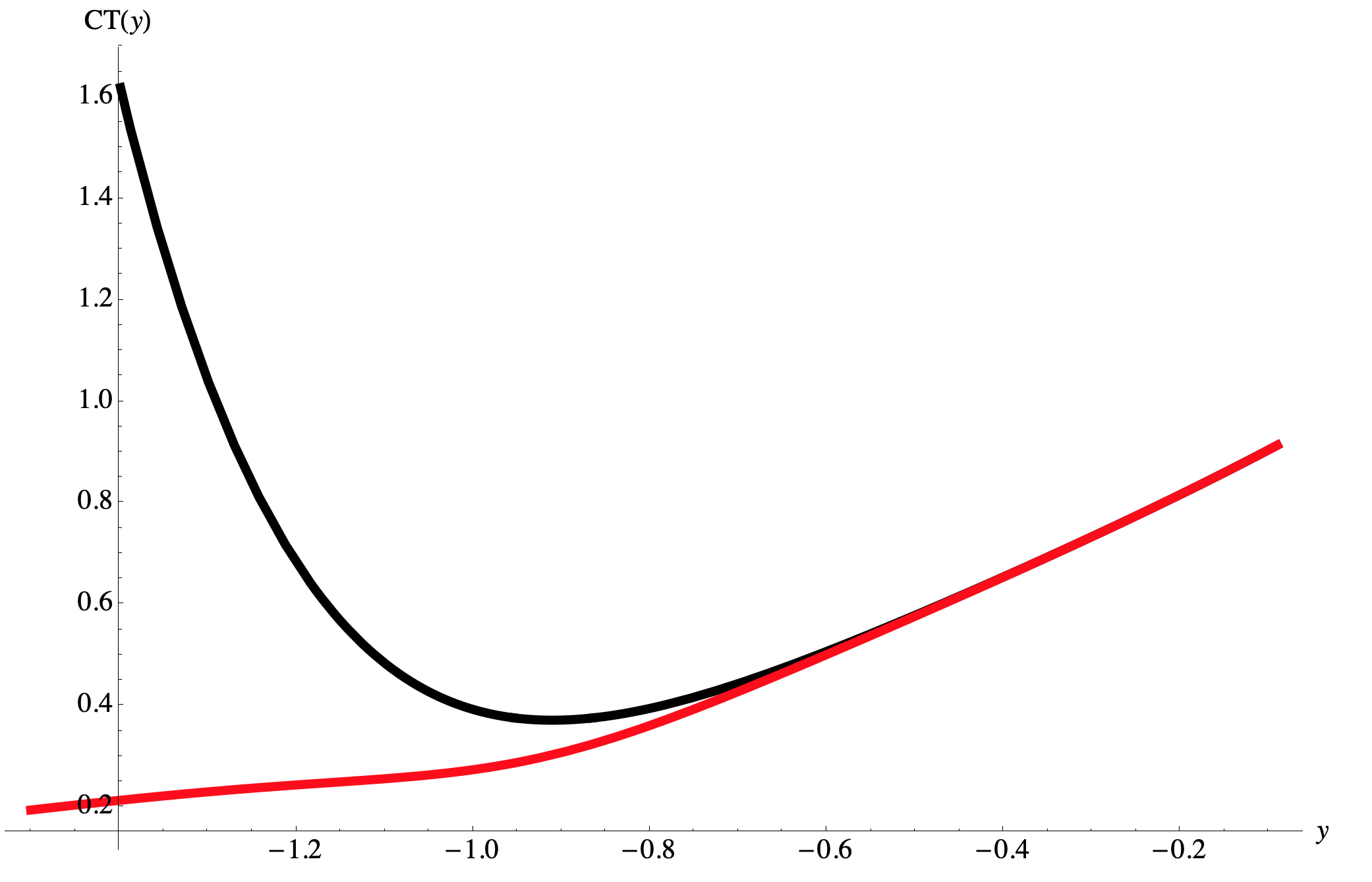}
   \epsfxsize=7.0cm
  \epsffile{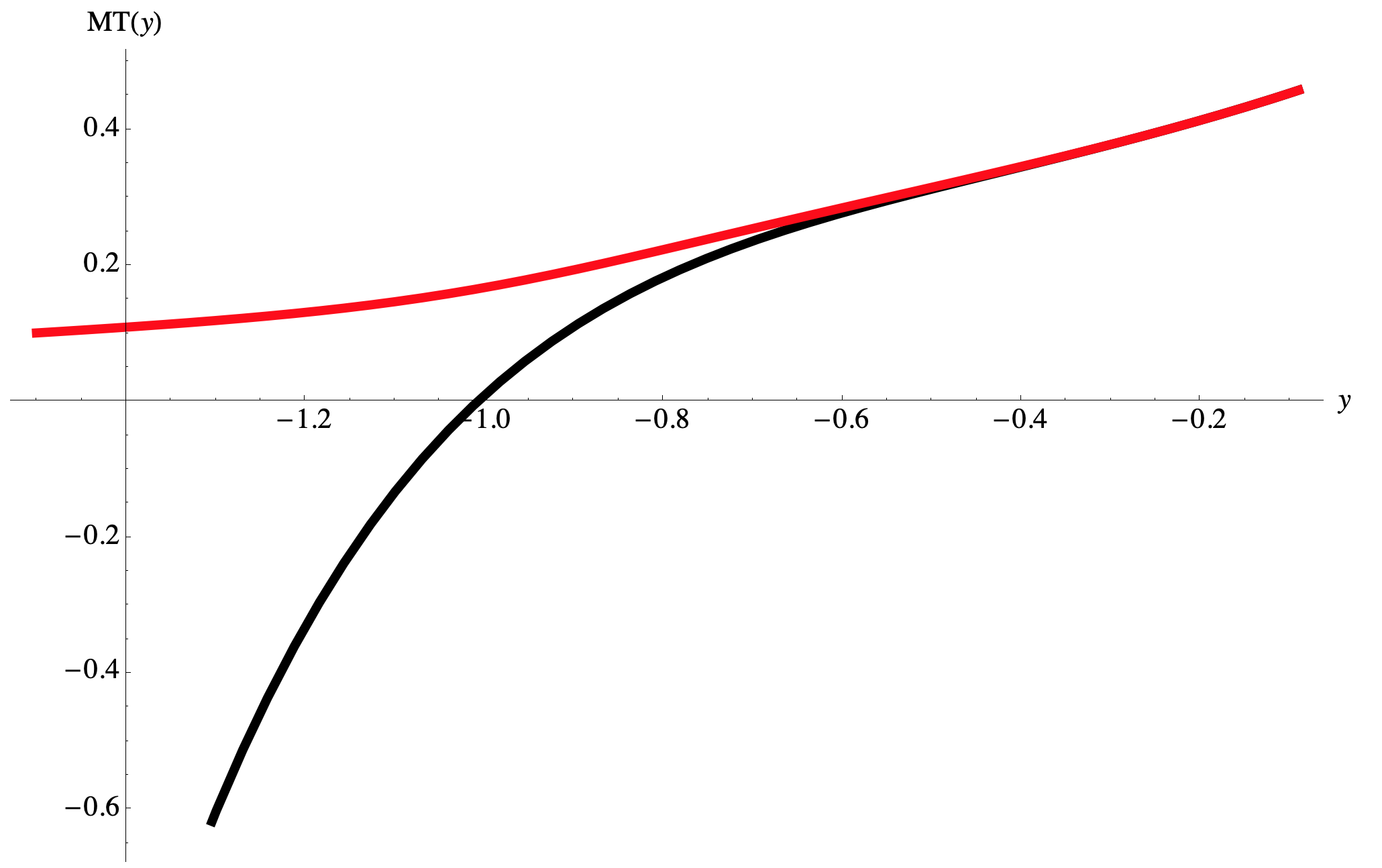}
  \end{center}\vspace{-0.2cm}
  \caption{The same as above but for the $CT$ and $MT$ 
  .}\label{PadeT}
  \end{figure}

To find the behaviour of the full amplitude one has to use numerical methods as was done for the case of supersymmetric theories in \cite{Borlakov:2016mwp,Kazakov:2016wrp,Borlakov:2017vra}. One can also use 
the Pad\'e approximation based on the first several terms generated by the recurrence relations (\ref{r1}), (\ref{r2}). 
 For example, the Pad\'e approximations $[3/3]$ for ${CS}(y)$ and ${MS}(y)$ based on the first 6 terms of expansion (\ref{CS},\ref{MS}) give:
\beqa
{CS}(y)&=&\frac{1 + \frac{919521}{17198}y + \frac{3619086}{214975}y^2 - 
 \frac{1132734289}{54173700}y^3}{1 + \frac{902323}{17198}y - 
 \frac{7767439}{214975}y^2 - \frac{34810827}{3009650}y^3}, \label{solPade} \\
  {MS}(y)&=&\frac{\frac 12 - \frac{60757261387}{27020023140}y - 
 \frac{17465208191899}{4458303818100}y^2 - 
 \frac{211448333535053}{1123492562161200}y^3}{
 1 - \frac{74267272957} {13510011570}y - \frac{7068734744869}{2229151909050}y^2 + 
 \frac{105130578087131}{16049893745160}y^3},
 \nonumber 
\eeqa
where we use the notation $y=sz^2/4$ and omitted the factor $z$ for $MS$. The corresponding plots as a function of $y$ are shown in Fig.\ref{PadeS}
One can repeat this analysis in the $t$-channel. The difference is that $t<0$ while $s>0$, which results in the change of the sign of $y$. The corresponding plots in the $t$ channel are shown in Fig.\ref{PadeT}.
Consideration of higher order Pad\'e approximations (up to $[9/9]$) dose not change the picture qualitatively in the $s$-channel, but gives more stable results in the $t$-channel. In the $s$-channel one always has a pole  located in the interval $(1.0,1.5)$, while the $t$-channel the pole for both $CT$ and $MT$ is absent. 

It is instructive to compare the Pad\'e approximations with the sequence of bubbles  which is summed up to the geometrical progression (\ref{sol}). The geometrical progression possesses the pole in the s-channel and smooth behaviour in the t-channel for both $CS$ and $MS$. One can see from the plots in Fig.\ref{PadeS} that the Pad\'e approximations also give the pole in the s-channel for both amplitudes and it is  quite stable. On the contrary, in the t-channel the $CT$ and $MT$ have a smooth behaviour for small values of $y$. Numerical analysis similar to the one used in \cite{Borlakov:2016mwp} also  suggests the presence of a pole in the $s$-channel.  For further analysis we assume that
the sequence of bubble diagrams captures the correct quantitative behaviour of the solution of equations (\ref{eq1}), (\ref{eq2}), that is the solution has a pole in the s-channel and no pole in the t-channel.

Remind that the pole terms described by $z=g/\epsilon$ are directly related to the leading log terms in the finite parts of the amplitudes. This allows one to study the high energy behaviour of the amplitudes when 
$s\sim t\sim u \sim E^2 \to \infty$. However, first one has to remove the UV divergences.  Not going into the discussion of this still unresolved problem we use the minimal subtraction scheme. As a partial justification of this procedure, we note that the leading poles as well as the leading logs do not depend on the subtraction scheme.

To get  the leading log term, one has to do  the replacement $z\to -g\log s$  in eq.(\ref{sol}) and similar for  other channels.
Then one has the  following asymptotics for the finite part of the amplitudes for the  sequence of  bubble diagrams:
\beqa\label{assympt}
&&CS=\frac{1}{1-g^2s \log^2 s/4},\ CT=\frac{1}{1-g^2t \log^2 t/4},\  CU=\frac{1}{1-g^2 u \log^2 u/4},\\
&&MS=-\frac{g\log s/2}{1-g^2s \log^2 s/4},\ MT=-\frac{g\log t/2}{1-g^2t\log^2 t/4},\  MU=-\frac{g\log u/2}{1-g^2\log^2 u/4}. \nonumber
\eeqa 
These expressions have a remarkable property. One has a pole in the s-channel and no poles in the t- and u-channels since $s>0$ and $t,u<0$. 

There can be several interpretations of such a pole. The first one can be made in the context of an effective theory. Here we link the coupling $g$ to the scale $E^{new}\sim 1/g$ when some new physics arises and trust our theory up to this scale $E^{new}$. Then the pole is located at the border of applicability of our effective theory and its presence can be considered as another indication that some "new physics" must appear at this scale. This implies that the logarithms we have summed up can be interpreted as IR logarithms, i.e. that  formulas (\ref{assympt}) describe the IR behaviour of the amplitude up to some scale $\sim 1/g$ \cite{Koschinski:2010mr,Linzen:2021fua}. This interpretation, as well as the structure of the expressions themselves, is close to the interpretation of the standard RG logarithms resummation in renormalizable theories within the context of critical behaviour~\cite{Vasilev:2004yr}. 

On the other hand, if we "take our theory seriously" \cite{Kazakov:2018zgi} and assume that there is a consistent way that allows one to remove UV divergences, then this pole can be interpreted as a new state in the spectrum.
This is exactly what happens when one has a resonance due to the exchange of a massive particle in the s-channel. In this case, the mass is equal to $1/g$ if one ignores the log  dependence compared to the power one. 
However, this state happens to be the ghost one since  the  sign of residue of the four-point amplitude is the same as
in the $\phi^4$ theory with  the Landau (ghost) pole.  

\section{Conclusion}
In this article, the analysis of the UV behaviour of the  Wess-Zumino model with the quartic interaction was performed.
We conclude 
that the structure of the UV divergences in this model
still allows quantitative analysis, despite this theory being  non-renorma\-lizable. One can calculate the UV divergences which follow the pattern of the general BPHZ-renormalization procedure. One can construct  recurrence relations which enable one to calculate all the leading divergences for the four point scattering amplitudes (and presumably any other observable) in a pure algebraic way starting from the one-loop ones. These recurrence relations can be further promoted to the generalized RG equations that have integro-differential form. These RG equations lead to the summation of the leading logs just like in renormalizable theories
and allow one to study the high energy behaviour of the scattering amplitudes and check the validity of the unitarity condition.

As it follows from our analysis, the WS model with quartic interaction essentially follows the pattern of the cubic Wess-Zumino model and has a ghost pole at high energy. Thus, it seems that without the gauge fields one is bounded to the Landau pole even in a non-renormalizable case.

We also found out that contrary to the renormalizable case in non-renormalizable theories the chiral part of the effective action obtains divergent radiative corrections. For the four-point amplitudes considered above they contain higher powers of momenta. However, due to the R-symmetry the structure of possible counterterms 
is restricted and the UV divergences are limited to the case of the chiral, anti-chiral and mixed amplitudes with four external fields. All the other amplitudes are globally UV convergent and may have only divergent subgraphs.

We have considered here the Wess-Zumino model since it  looks relatively simple and is an example of non-renormalizable model in 4 dimensions. However, our approach has general validity and can be applied  to any model. 
It would be interesting to apply it to the analysis of the UV structure of gravity theories with or without supersymmetry. We are going to discuss it in more detail in upcoming publications.

\section*{Acknowlegments}
The authors is grateful to  J.Buchbinder and K.Stepanyantz for useful discussions and check of the calculations of the diagrams.  We thank E.Ivanov who pointed out the importance of R-symmetry in the present analysis. The participation of D.Tolkachev and R.Iakhib\-baev at the early stage of this work is cordially acknowledged. The work of DK was supported by the Russian Science Foundation grant \# 21-12-00129 and the work of LB by the Basis Foundation.

\appendix

\section{Appendix}
\label{ap1}

In this appendix, we briefly discuss our approach to the $N=1$ superspace computations and also present a complete list of superspace diagrams and their divergent parts which we used in the main text.

To compute  the functions $C$ and $M$ one can note that the standard Feynman diagrams contributing to
the four-point scattering amplitudes are identical to those contributing to the strongly connected correlation functions if the on-shell condition and multiplication by the corresponding polarisation vectors for external lines are taken into account. 

The strongly connected correlation functions can be extracted form the effective action $\Gamma[\Phi,\bar \Phi]$ which in turn can be computed perturbatively using the $N=1$ superspace Feynman diagrams \cite{Gates:1983nr}. The Feynman rules for our theory are condensed in Fig.\ref{figFRules}. They are obviously very similar to
the ordinary WZ model, the only difference is that now we have quartic vertex instead of triple one that  generates additional $D^2$ and $\bar D^2$ factors.

The relevant parts of the effective action $\Gamma$ at $l$ loops for $\langle \Phi_1 \Phi_2 \Phi_3 \Phi_4 \rangle$ and $\langle \Phi_1 \Phi_2 \bar\Phi_3 \bar\Phi_4 \rangle$, which we label as $\Gamma_4[\Phi]$ and $\Gamma_4[\Phi,\bar \Phi]$, can be written as
\begin{eqnarray}
\Gamma_4^{(l)}[\Phi]=\int\prod_{m=1}^4d^4p_m d^4\theta_m ~\Phi(p_m,\theta_m)~
\sum_i G_i,
\end{eqnarray}
and
\begin{eqnarray}
\Gamma_4[\Phi,\Bar \Phi]=\int\prod_{i=m}^4d^4p_m d^4\theta_m ~\Phi_1(\theta_1)\Phi_2(\theta_2) 
\bar \Phi_3(\bar\theta_3) \bar \Phi_4(\bar\theta_4)~
\sum_iG_i,
\end{eqnarray}
respectively. Here $G_i$ are the $l$-loop supergraphs contributing to the chiral and mixed correlation functions.  All the relevant supergraphs up to $l=4$ are presented below in Table \ref{tab1} and \ref{tab2}. 

Using the standard $D$-algebra described in classical textbooks \cite{Gates:1983nr,West:1990tg}, one can simplify these expressions and impose the on-shell constraints on external legs. We put all $p_i^2=0$ for external momenta and impose the equation of motion constraints on 
$\Phi$ and $\bar \Phi$, namely 
$\bar D^2 \bar \Phi=0$, $D^2 \Phi=0$.
This results in
\begin{eqnarray}
\Gamma_4^{(l)}[\Phi]=\int d^2\theta~\prod_{m=1}^4d^4p_m\Phi_m(\theta)~
\sum_i I_i,
\end{eqnarray}
and
\begin{eqnarray}
\Gamma^{(l)}_4[\Phi,\bar \Phi]=\int d^4\theta ~\prod_{i=1}^4d^4p_i ~\Phi_1(\theta)\Phi_2(\theta) 
\bar \Phi_3(\bar\theta) \bar \Phi_4(\bar\theta)~
\sum_i I_i,
\end{eqnarray}
where $ I_i$ are the ordinary Feynman $l$-loop scalar integrals shown in Table \ref{tab1} and \ref{tab2}. 

\begin{table}[!h]
\setlength\arrayrulewidth{0.7pt}
\begin{tabular}{|l|l|c|c|}
\hline
 \multicolumn{1}{|c|}{\small{Super Diagram $G_i$}} & \multicolumn{1}{|c|}{\small{Scalar Diagram $I_i$}} & \multicolumn{1}{|c|}{\small{Highest Pole}} & \multicolumn{1}{|c|}{\small{Comb.}} \\ 
 \hline
 \parbox{1cm}{\includegraphics[scale=0.06]{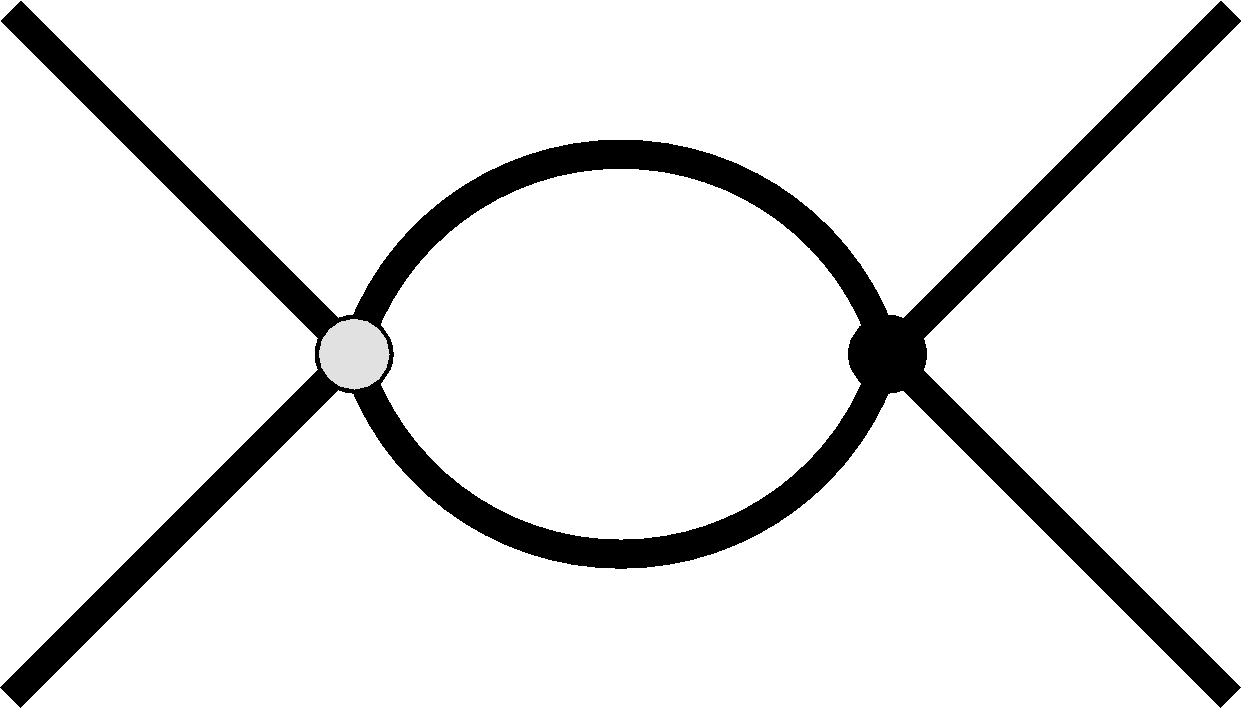}} & 
 \parbox{1cm}{\includegraphics[scale=0.05]{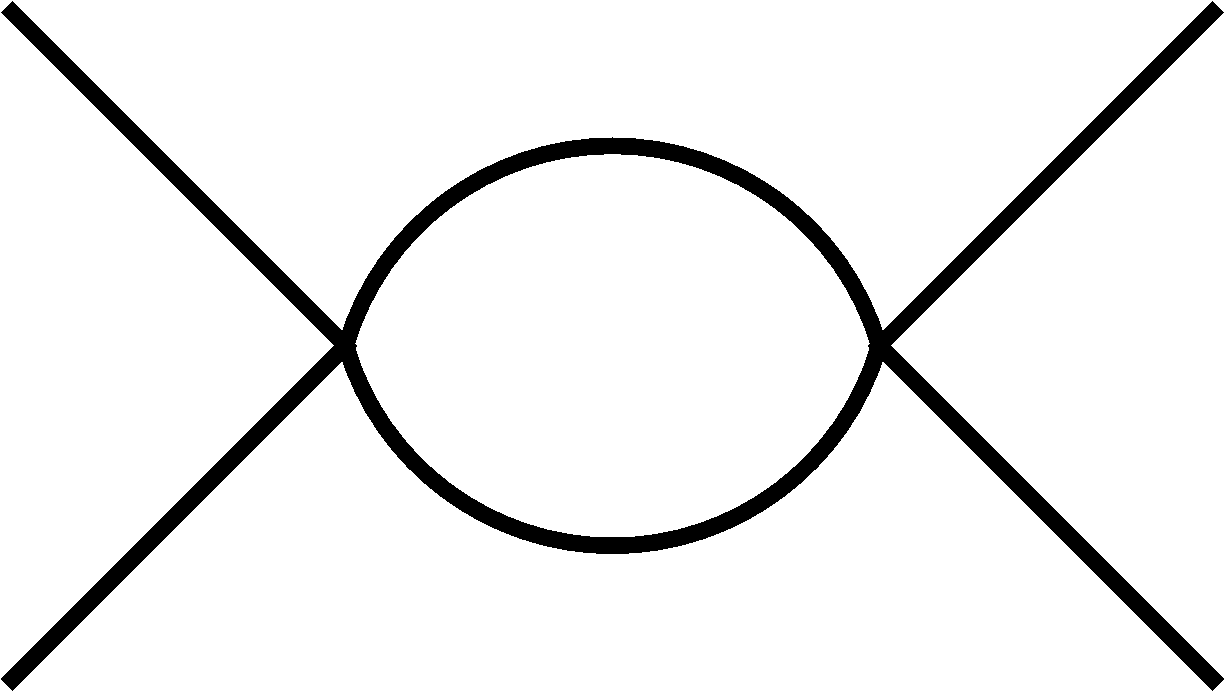}} & 
 \parbox{1cm}{
 \begin{align}
     \frac{1}{\epsilon}\nonumber
 \end{align}} &
 \parbox{1cm}{
 \begin{align}
      \frac{1}{2}\nonumber
  \end{align}} \\ 
\hline
 \parbox{4cm}{\includegraphics[scale=0.06]{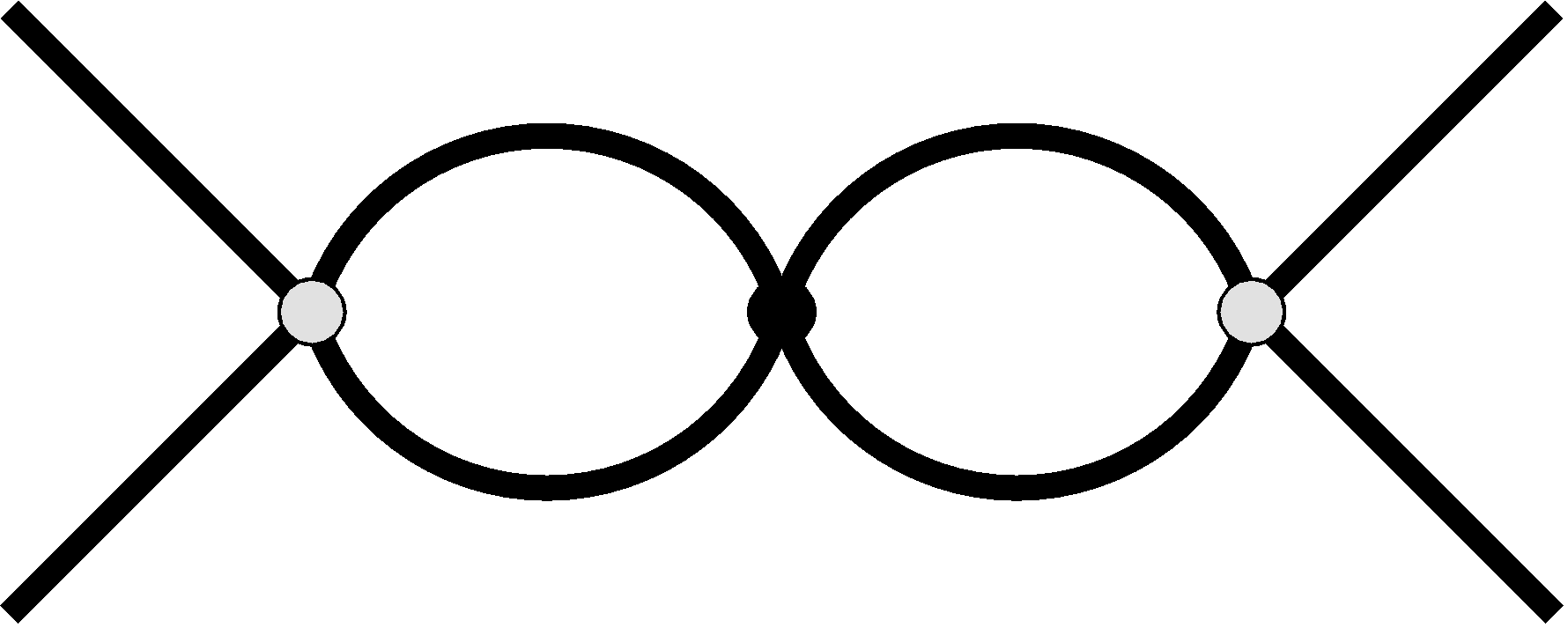}} &  
 \parbox{4cm}{\includegraphics[scale=0.05]{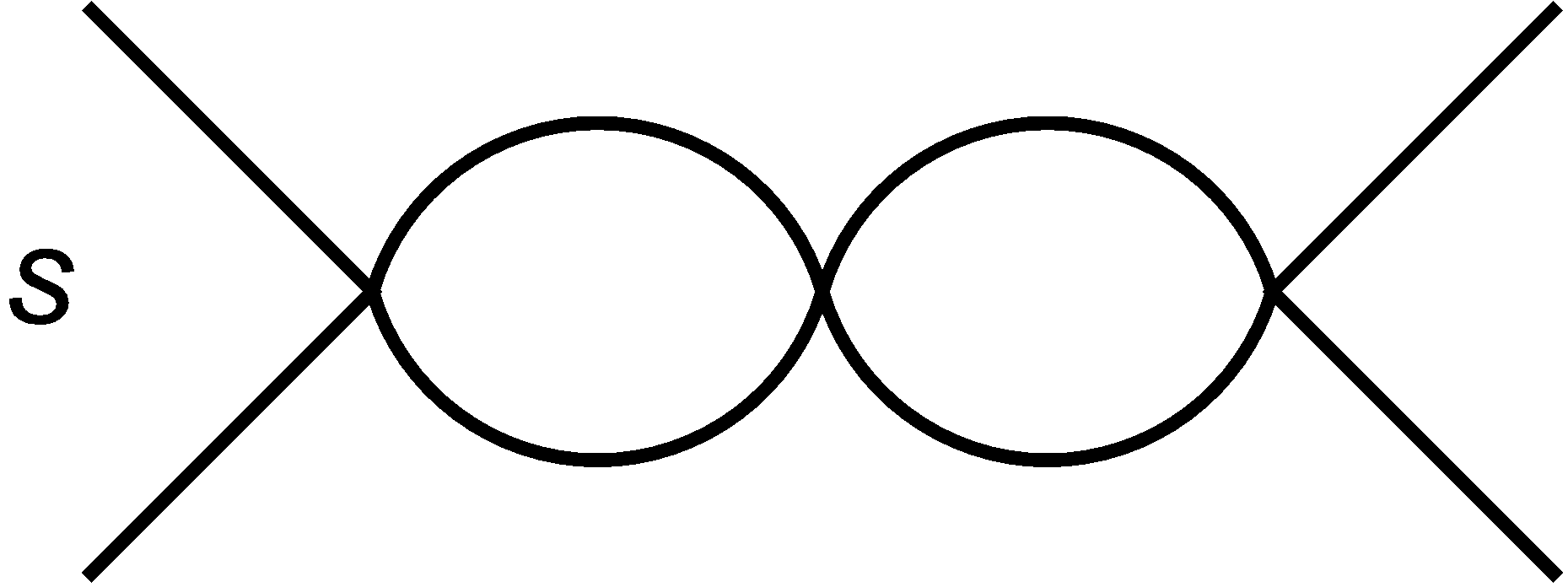}} &  
\parbox{1cm}{
 \begin{align}
     \frac{s}{\epsilon^2}\nonumber
 \end{align}} &  
\parbox{1cm}{
 \begin{align}
     \frac{1}{4}\nonumber
 \end{align}}\\ 
 \hline
 \parbox{5cm}{\includegraphics[scale=0.06]{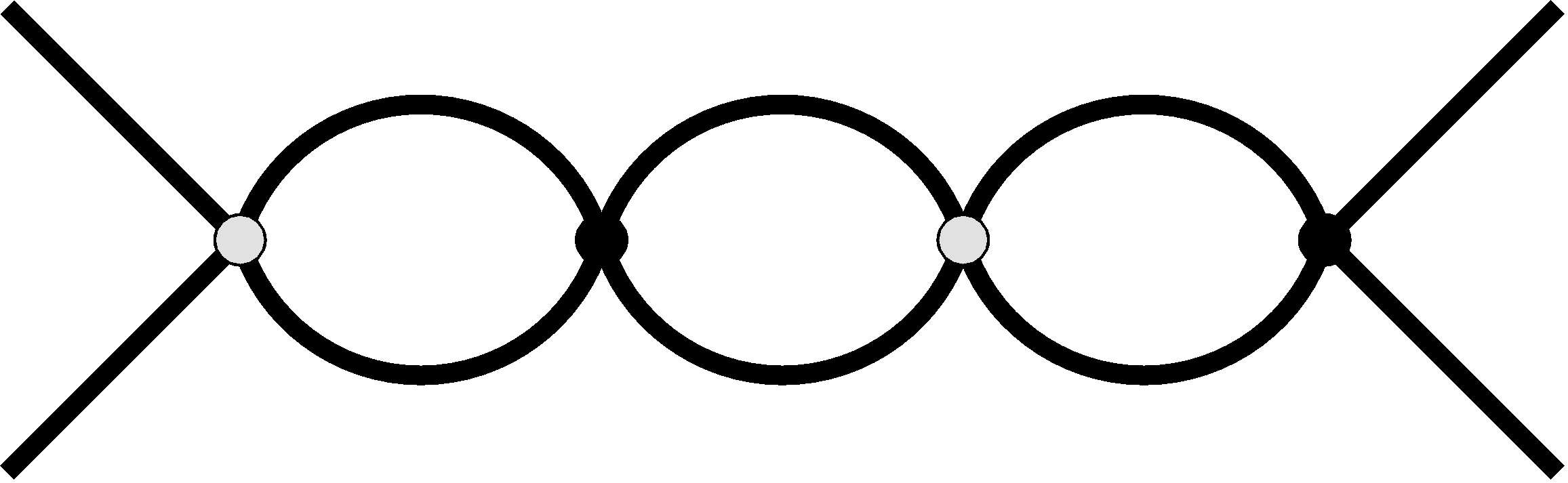}} & 
 \parbox{5cm}{\includegraphics[scale=0.05]{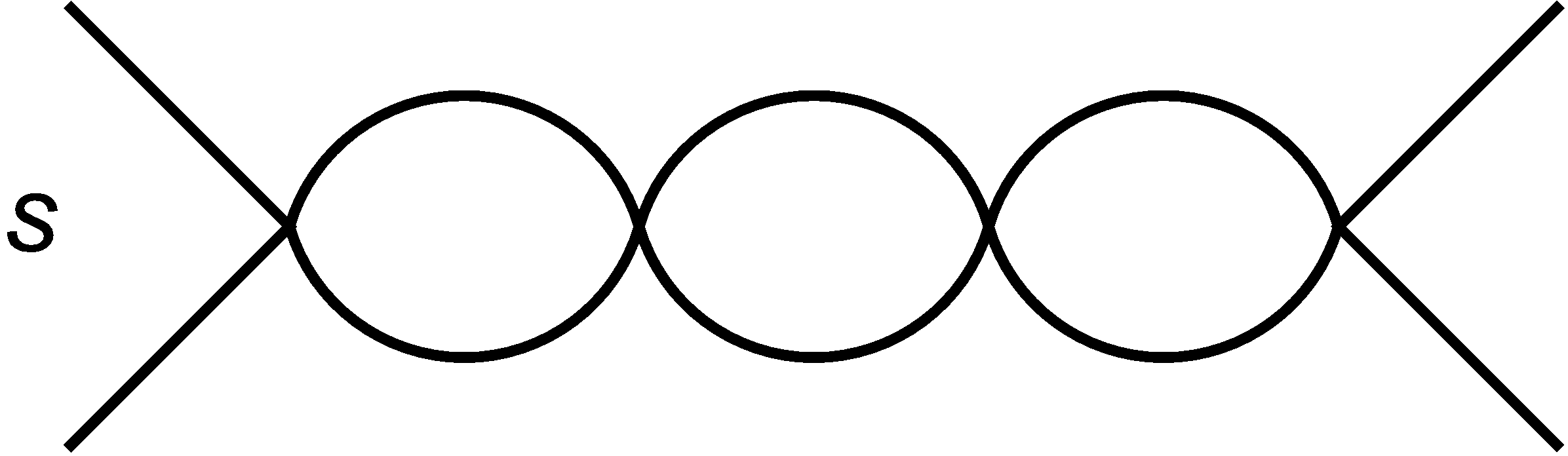}} &  
 \parbox{1cm}{
  \begin{align}
      \frac{s}{\epsilon^3}\nonumber
  \end{align}} &  
 \parbox{1cm}{
  \begin{align}
      \frac{1}{8}\nonumber
  \end{align}} \\ 
\hline
 \parbox{5cm}{\includegraphics[scale=0.06]{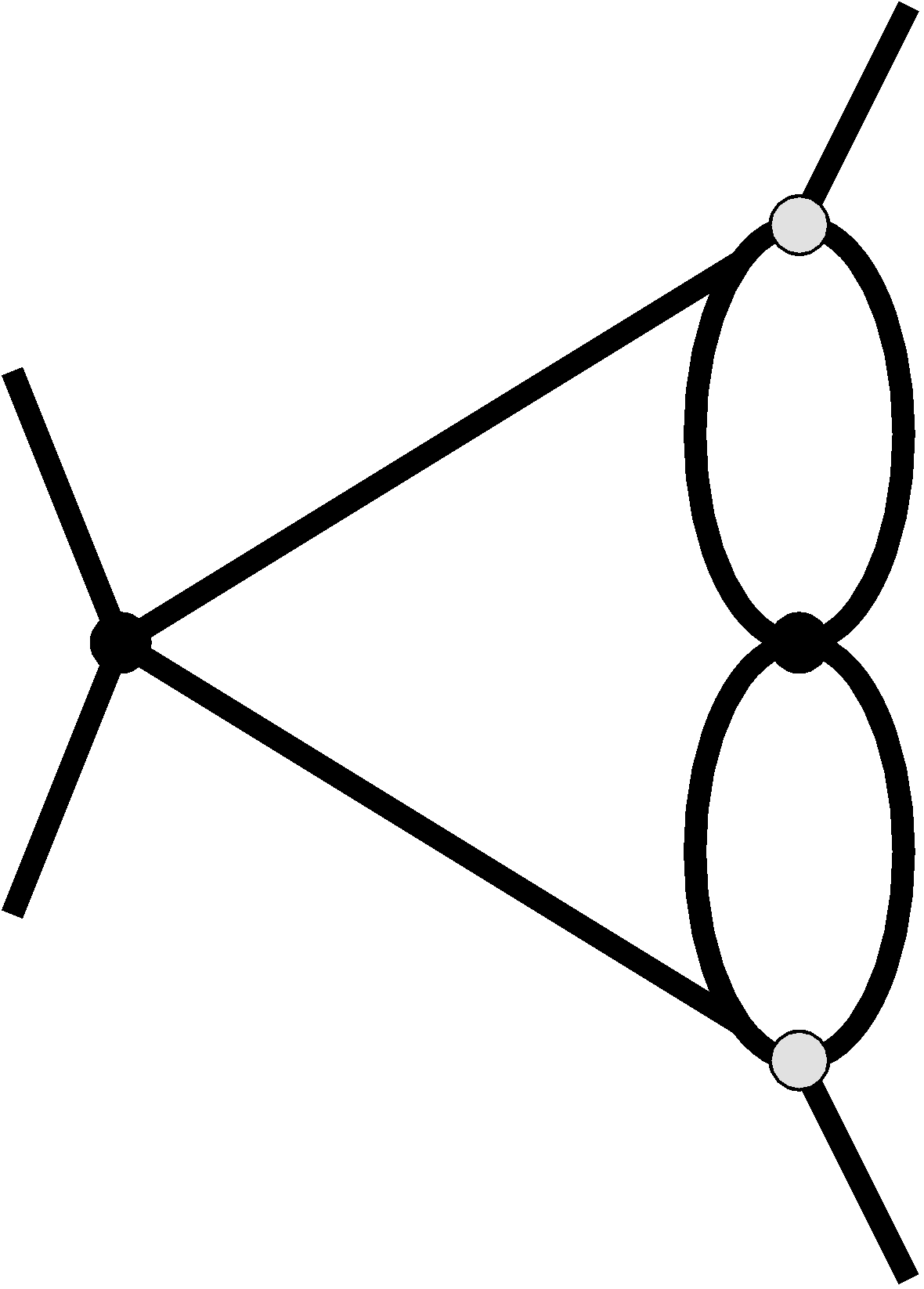}} & 
 \parbox{5cm}{\includegraphics[scale=0.05]{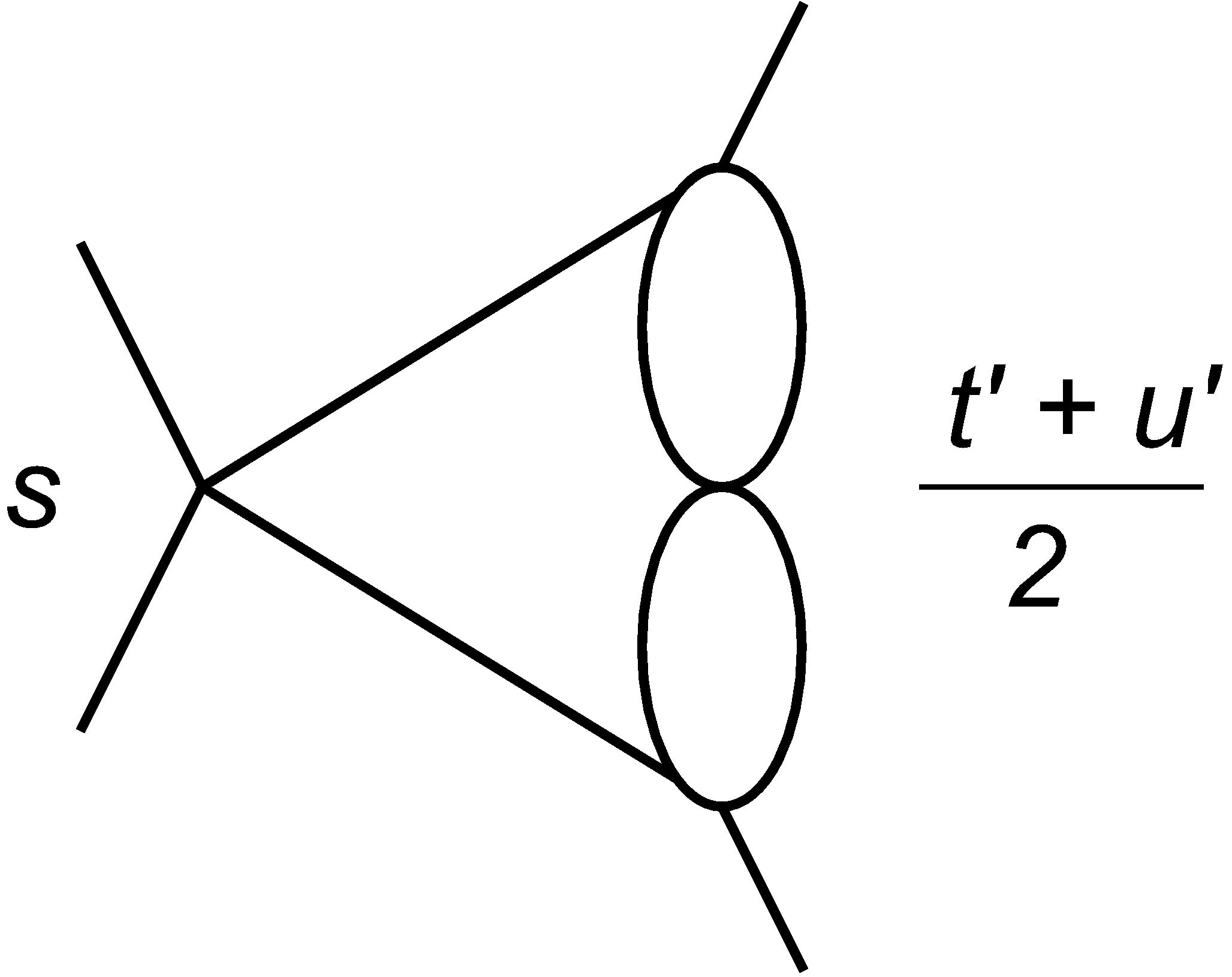}} &  
 \parbox{1cm}{
   \begin{align}
      -\frac{1}{2}~\frac{s}{3\epsilon^3}\nonumber
   \end{align}} &  
 \parbox{1cm}{
    \begin{align}
       2\times\frac{1}{2}\nonumber
    \end{align}} \\ 
 \hline
 \parbox{5cm}{\includegraphics[scale=0.06]{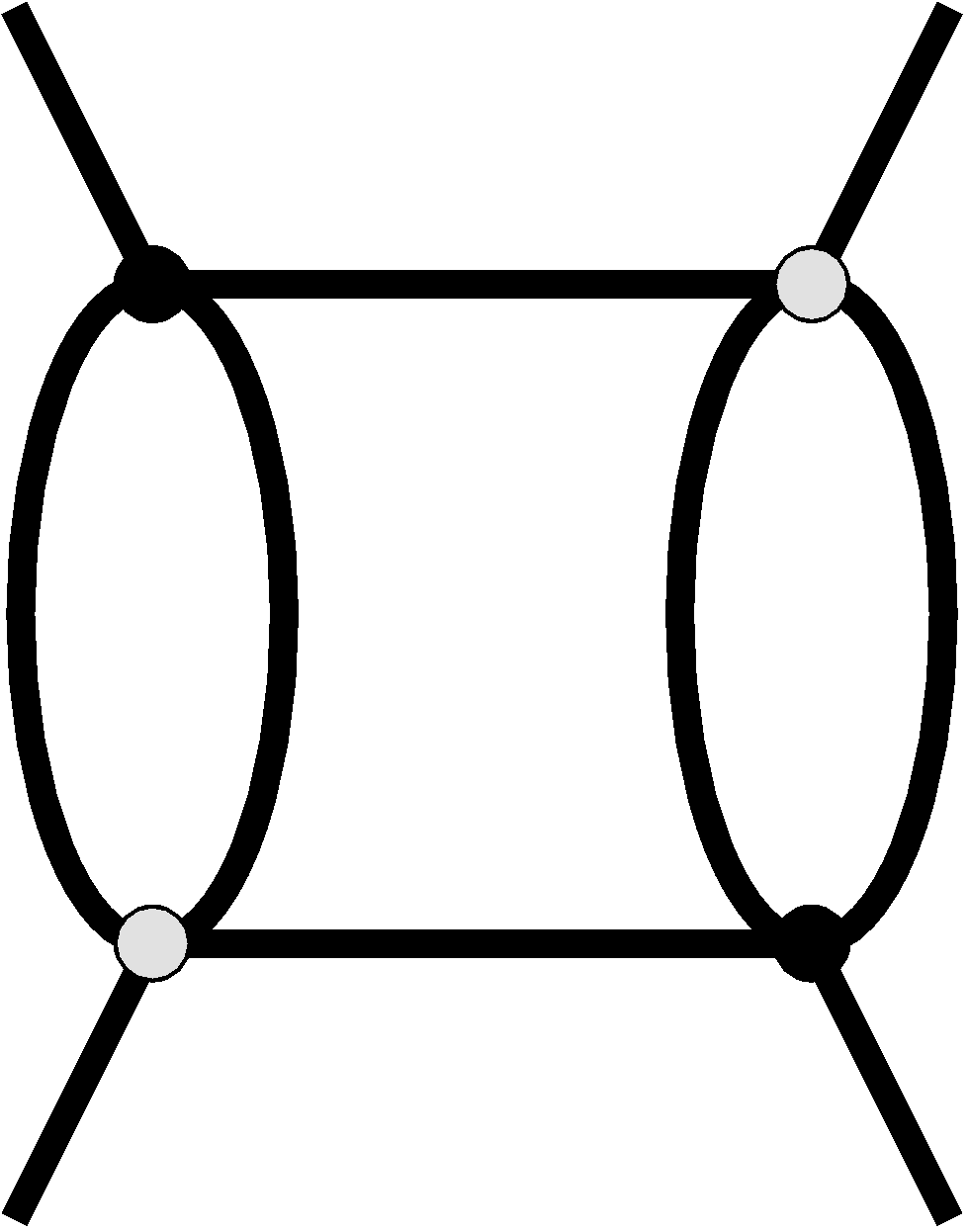}} & 
 \parbox{5cm}{\includegraphics[scale=0.05]{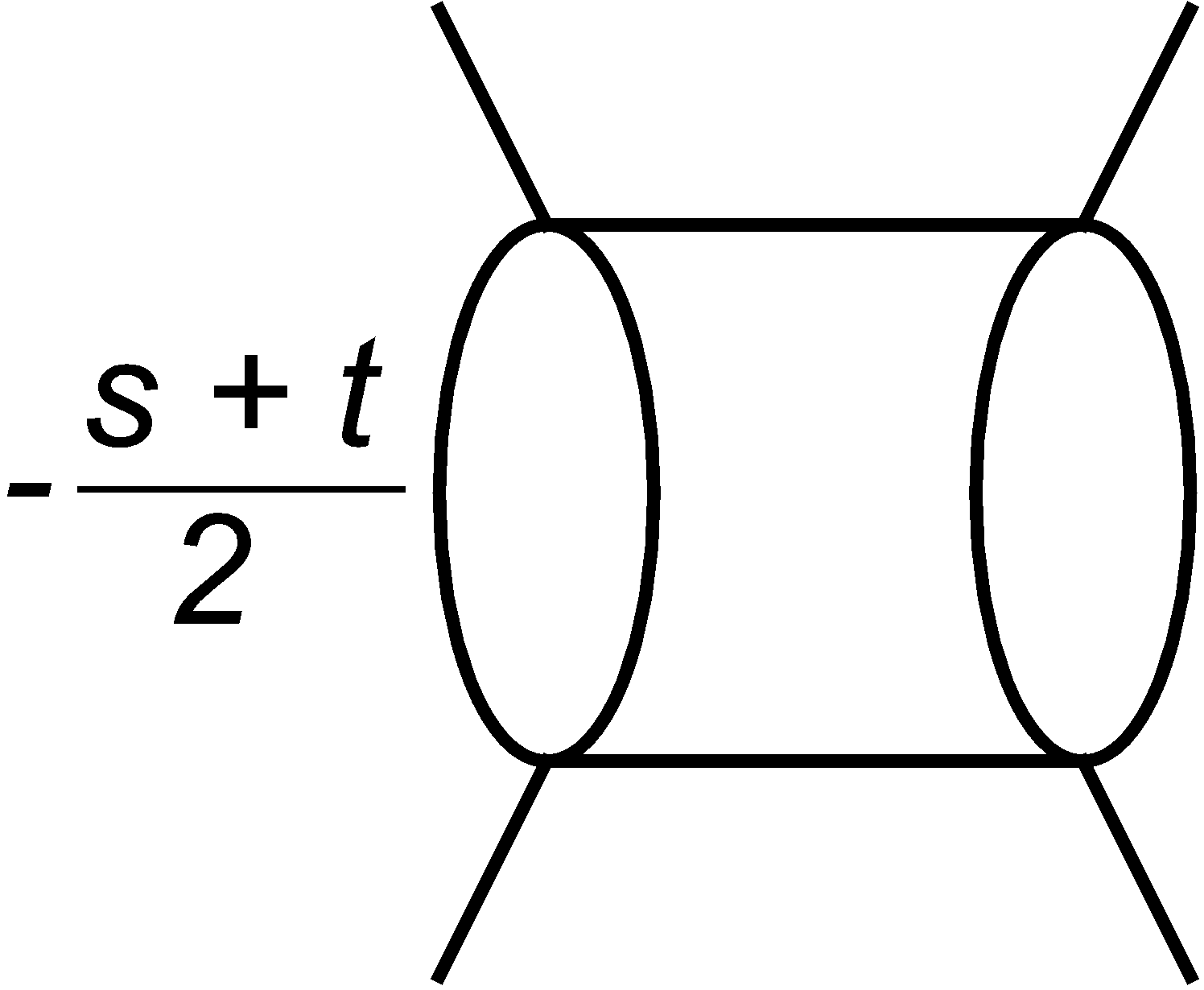}} &  
 \parbox{1cm}{\begin{align}
       +\frac{1}{2}~\frac{s}{3\epsilon^3}\nonumber
    \end{align}}&  
 \parbox{1cm}{\begin{align}
        2\times\frac{1}{2}\nonumber
     \end{align}}\\ \hline
\end{tabular}
\caption{Feynman diagrams contributing to the four-point scattering amplitude in the lower loops  in the s-channel (leading divergences only). Part one. The white (black) dot denotes the chiral (anti-chiral) vertex, $D$ factors are not shown. In the second column the corresponding scalar diagrams obtained after performing the Grassmannian algebra are shown. The third and fourth columns contain the leading poles and combinatoric factors \label{tab1}}
\end{table}

\begin{table}[!h]
\setlength\arrayrulewidth{0.7pt}
\begin{tabular}{|l|l|c|c|}
\hline
 \multicolumn{1}{|c|}{\small{Super Diagram $G_i$}} & \multicolumn{1}{|c|}{Scalar Diagram $I_i$} & \multicolumn{1}{|c|}{Highest Pole} & \multicolumn{1}{|c|}{Comb.} \\ 
\hline
 \parbox{6cm}{\includegraphics[scale=0.06]{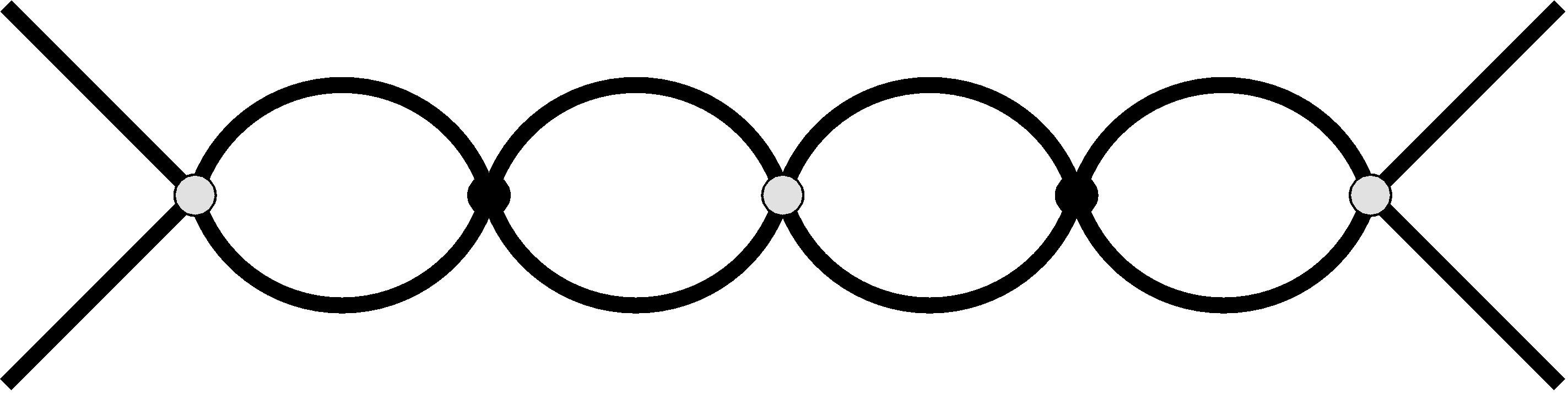}} & 
 \parbox{5.5cm}{\includegraphics[scale=0.05]{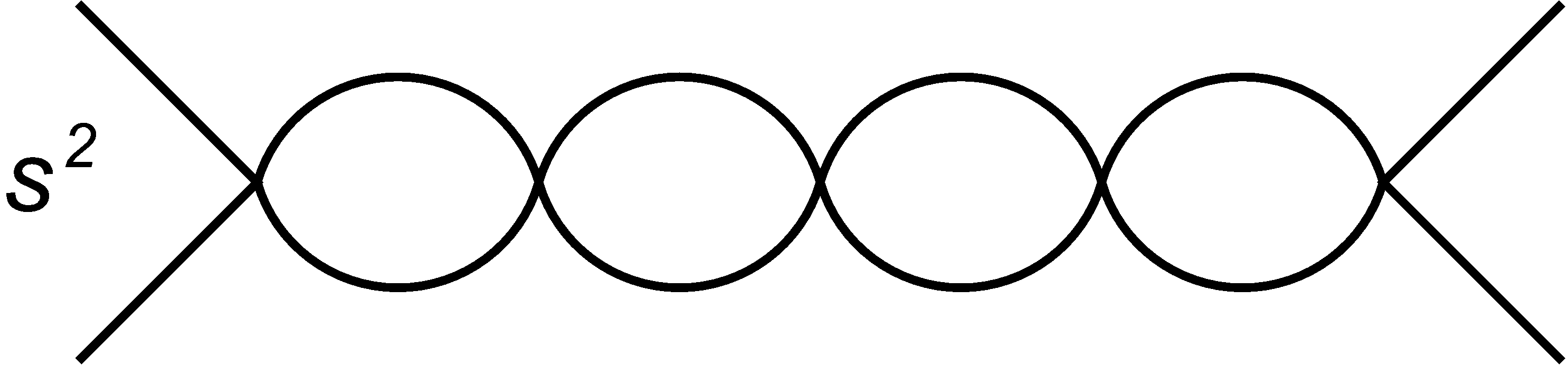}} &  
  \parbox{1cm}{\begin{align}
        \frac{s^2}{\epsilon^4}\nonumber
     \end{align}}&  
  \parbox{1cm}{\begin{align}
         \frac{1}{16}\nonumber
      \end{align}}\\ 
 \hline
 \parbox{6cm}{\includegraphics[scale=0.06]{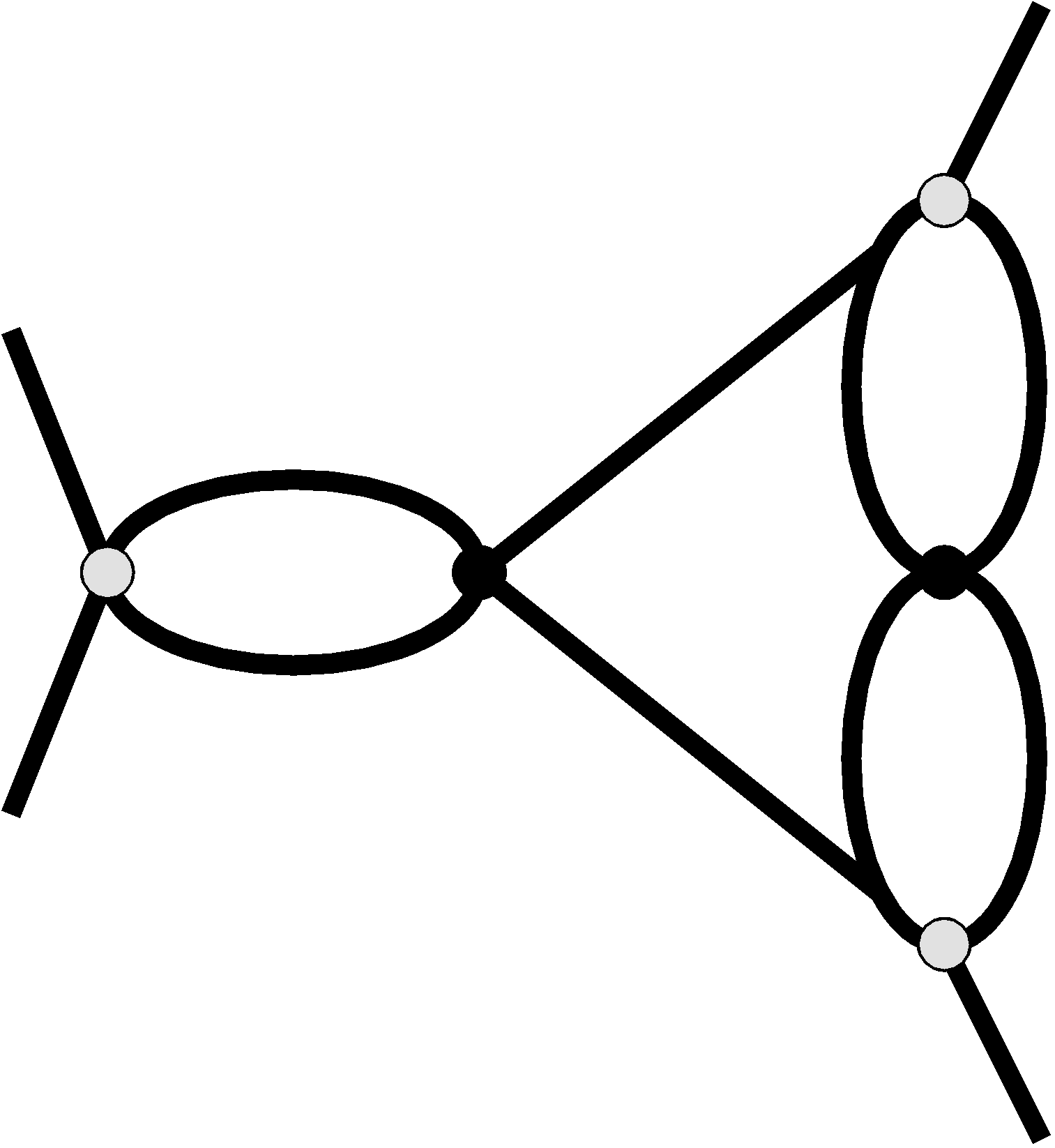}} & 
 \parbox{5.5cm}{\includegraphics[scale=0.05]{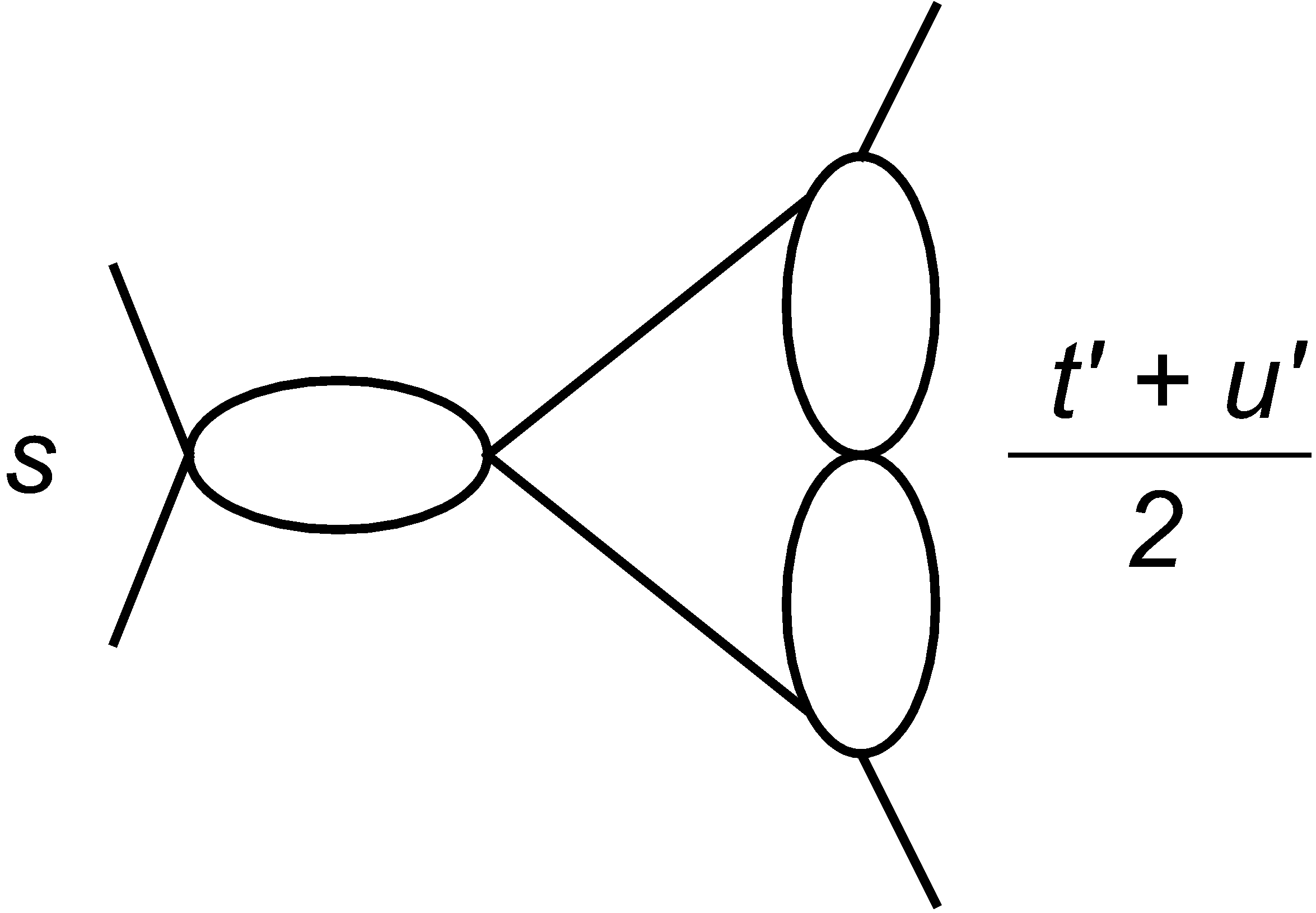}} &  
  \parbox{1cm}{\begin{align}
        -\frac{1}{2}~\frac{s^2}{3\epsilon^4}\nonumber
     \end{align}}&  
  \parbox{1cm}{\begin{align}
         \frac{1}{4}\nonumber
      \end{align}}\\ 
 \hline
 \parbox{6cm}{\includegraphics[scale=0.06]{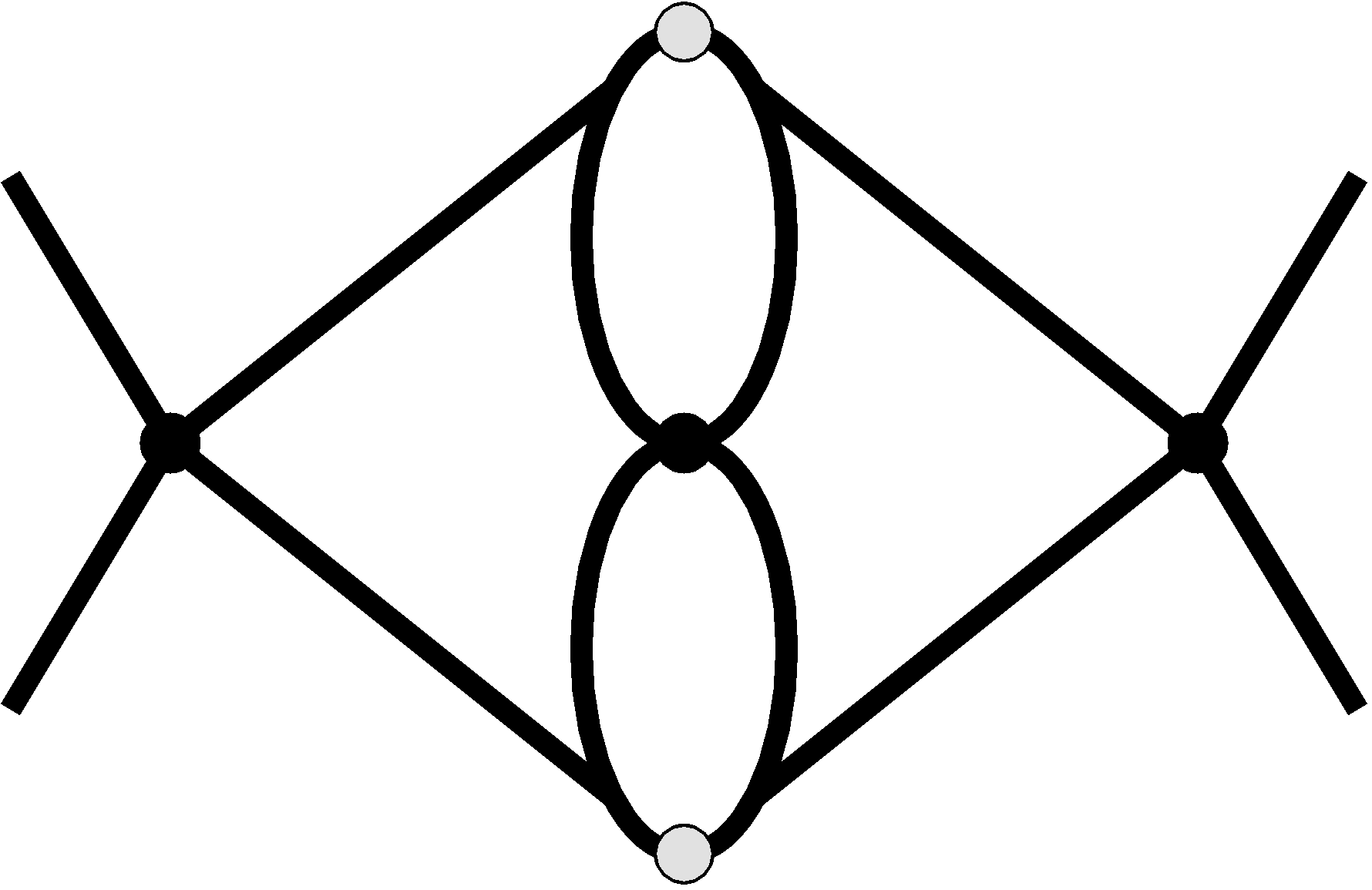}} & 
 \parbox{5.5cm}{\includegraphics[scale=0.05]{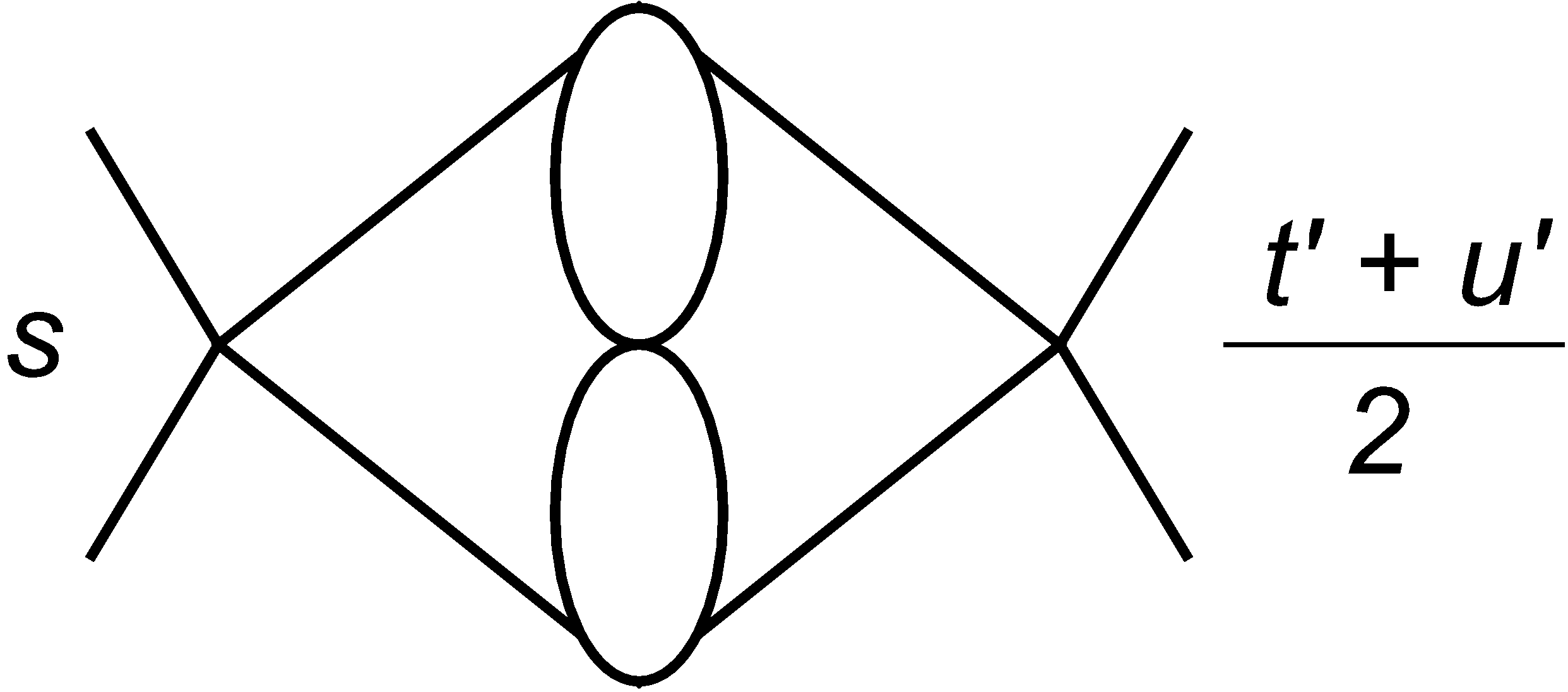}} &  
  \parbox{1cm}{\begin{align}
        -\frac{1}{2}~\frac{s^2}{6\epsilon^4}\nonumber
     \end{align}}&  
  \parbox{1cm}{\begin{align}
         \frac{1}{8}\nonumber
      \end{align}}\\ 
 \hline
 \parbox{6cm}{\includegraphics[scale=0.06]{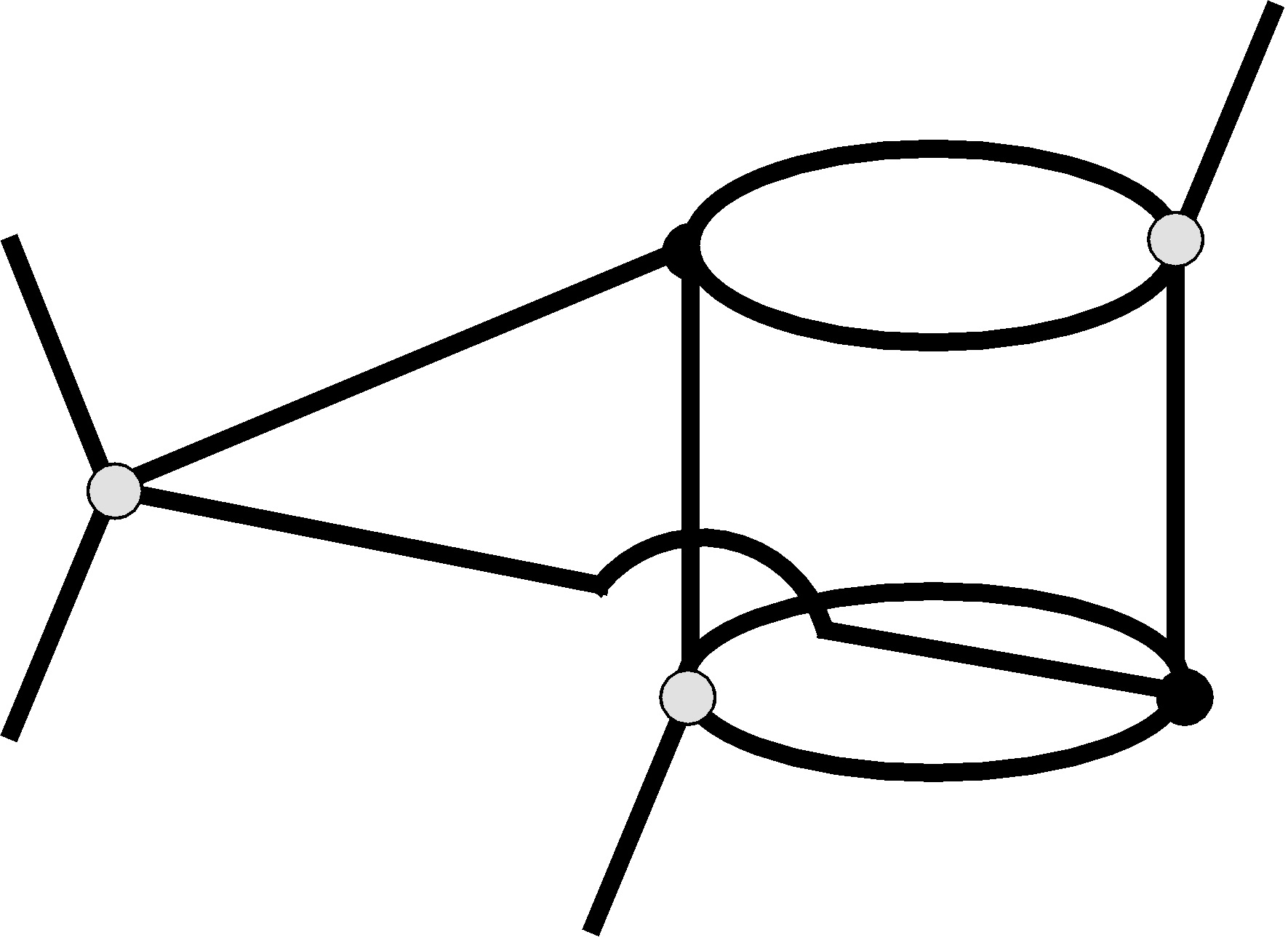}} & 
 \parbox{5.5cm}{\includegraphics[scale=0.05]{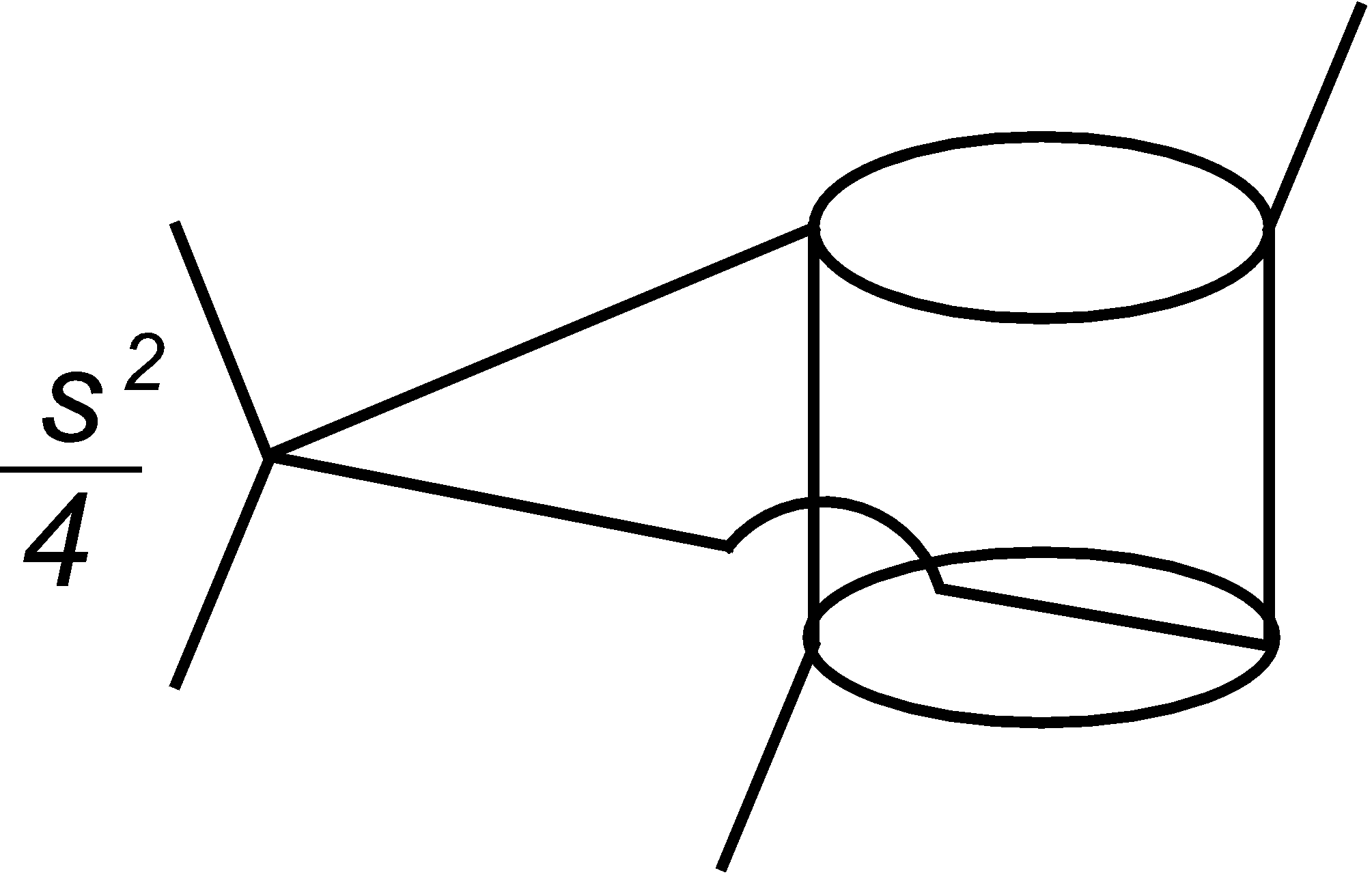}} &  
  \parbox{1cm}{\begin{align}
        +\frac{1}{4}~\frac{s^2}{12\epsilon^4}\nonumber
     \end{align}}&  
  \parbox{1cm}{\begin{align}
         1\nonumber
      \end{align}}\\ \hline
\end{tabular}
\caption{Feyman diagrams contributing to the four-point scattering amplitude in the lower loops  in the s-channel (leading divergences only). Part two. The white (black) dot denotes the chiral (anti-chiral) vertex, $D$ factors are not shown. In the second column the corresponding scalar diagrams obtained after performing the Grassmannian algebra are shown. The third and fourth columns contain the leading poles and combinatoric factors \label{tab2}}
\end{table}
Note that all these integrals are given by a subset of Feynman integrals of the
scalar $\phi^4$ theory with some numerators. The reduction of supergraphs to the usual graphs is performed by the standard $N=1$ superspace technique. The only relatively complicated case is the last non-planar supergraph in Table \ref{tab2}. In this case, after the $D$-algebra one obtains:  
\begin{eqnarray}
\frac{1}{4}\int\prod_{m=1}^4d^4p_m d^4\theta~F^{\alpha\beta} 
\int \frac{d^Dl~d^Dq~d^Dx~d^Dy \left( \epsilon^{\dot \gamma \dot \delta} (l)_{\alpha \dot\delta}(k)_{\beta \dot \gamma}-2\epsilon_{\gamma\alpha}(k)_{\beta\dot \delta}+\epsilon_{\gamma\beta}(l)_{\alpha\dot \delta}(k)^{\gamma\dot \delta}\right)}{l^2(l-p_{12})^2q^2(q-l-p_4)^2x^2(q-l+p_{12}-x)^2y^2(q+p_4-y)^2},\label{comp}
\end{eqnarray}
where ($k\equiv l+p_4-q$)
and
\begin{eqnarray}
F^{\alpha\beta}=[D^{\alpha}(\Phi_1(\theta)\Phi_2(\theta))]~[D^{\beta}\Phi_3(\theta)]~\Phi_4(\theta).
\end{eqnarray}
The numerator in (\ref{comp}) originates from the integration by parts and further simplifications of monomials of the $D$ derivatives.

Then, after replacement $d^4\theta = d^2\theta \bar D^2$ this expression can be simplified to:
\begin{eqnarray}
\int d^2\theta~\prod_{m=1}^4d^4p_m\Phi_m(\theta) ~I[\mbox{tr}_+(lkp_4p_3)],
\end{eqnarray}
where:
\begin{eqnarray}
I[\mbox{tr}_+(lkp_4p_3)]=\int \frac{d^Dl~d^Dq~d^Dx~d^Dy~\mbox{tr}_+(lkp_4p_3)}{l^2(l-p_{12})^2q^2(q-l-p_4)^2x^2(q-l+p_{12}-x)^2y^2(q+p_4-y)^2}.
\end{eqnarray}
The $\mbox{tr}_+$ corresponds to the standard trace with $(1+\gamma_5)$ insertion which will generate additional term proportional to the contraction of momenta with $\epsilon$-symbol. This term however
will drop out in our case due to the momentum conservation delta function constraint. By carefully analysing three remaining terms, it can be shown that the leading divergent part of the $I[\mbox{tr}_+(lkp_4p_3)]$ scalar integral is given by:
\begin{eqnarray}
I[\mbox{tr}_+(lkp_4p_3)]\Big|_{div.}=\frac{s^2}{4}I[1]\Big|_{div.},
\end{eqnarray}
where the integral $I[1]$ corresponds to the last four-loop scalar none-planar diagram in Table \ref{tab2}.

\section{Appendix}
\label{ap2}
In this appendix, we briefly discuss the structure of the scattering amplitudes in the WZ theory in on-shell momentum superspace \cite{Henrietta_Amplitudes,Elvang:2011fx} which is usually used in modern approaches to scattering amplitude computations. This allows us to obtain simple expressions for the polarisation factors from (\ref{schemAmpl}) in terms of the Grassmannian delta functions.

In the case when CPT-conjugated supermultiplets are present one can use the non-chiral version of the on-shell momentum superspace (the $\Psi$-$\Psi^{\dagger}$ formalism in the terminology of \cite{Elvang:2011fx}) which can be parametrised by
\begin{equation}
\{\lambda^{\alpha},\tilde{\lambda}^{\dot{\alpha}},\eta,\bar{\eta}\},
\end{equation}
where $\lambda^{\alpha}$ and $\tilde{\lambda}^{\dot{\alpha}}$ are the bosonic $SL(2,C)$ spinors which originate from the  on-shell momenta of massless
particles $p^{\alpha\dot{\alpha}}=\lambda^{\alpha}\tilde{\lambda}^{\dot{\alpha}}$, and $\eta$,$\bar{\eta}$
are the fermionic scalars.

All the creation/annihilation operators of the on-hell states of the theory (\ref{LWZ}) can be combined into two superfields $\Omega_i$ and $\bar{\Omega}_i$:
\begin{eqnarray}
\Omega_i&=&a^{(0)}(p_i)+a^{(+\frac{1}{2})}(p_i)\eta_i,\nonumber\\
\bar{\Omega}_i&=&a^{*(0)}(p_i)+a^{(-\frac{1}{2})}(p_i)\bar{\eta}_i,
\end{eqnarray}
where $a^{(0)},a^{*(0)}$ correspond to the creation/annihilation operators of scalars and $a^{(\pm\frac{1}{2})}$ correspond to the creation/annihilation operators of fermions.

The supertranslation generators for the $n$-particle case are realised in this superspace as:
\begin{eqnarray}
p^{\alpha \dot \beta}&=&\sum_{i=1}^n \lambda_i^{\alpha} \tilde \lambda_i^{\dot \beta},\nonumber\\
q^{\alpha}&=&\sum_{i=1}^n \lambda_i^{\alpha}\eta_i+\lambda^{\alpha}_i\frac{\partial}{\partial \bar \eta_i},\nonumber\\
\bar q^{\dot \alpha}&=&\sum_{i=1}^n \tilde \lambda_i^{\dot \alpha}\bar\eta_i+\tilde \lambda^{\dot \alpha}_i\frac{\partial}{\partial \eta_i}.
\end{eqnarray}
Using the $\Omega$ and $\bar \Omega$ superfields all
the nonvanishing four-point scattering amplitudes in this theory can be 
combined into the three super-amplitudes defined on the on-shell momentum superspace:
\begin{equation}
A_4(\Omega_1,\Omega_2,\Omega_3,\Omega_4),A_4(\bar{\Omega}_1,\bar{\Omega}_2,\bar{\Omega}_3,\bar{\Omega}_4)
,A_4(\Omega_1,\Omega_2,\bar{\Omega}_3,\bar{\Omega}_4).
\end{equation}
The correspondence between the Grassmannian variables and the on-shell states are given by
the following relations for the chiral
\begin{eqnarray}
a^{(0)}(p_i) &\leftrightarrow & \eta_i^0~~~a^{(+\frac{1}{2})}(p_i)\leftrightarrow \eta_i^1
\end{eqnarray}
anti-chiral,
\begin{eqnarray}
a^{*(0)}(p_i) &\leftrightarrow & \bar\eta_i^0~~~a^{(-\frac{1}{2})}(p_i)\leftrightarrow \bar\eta_i^1
\end{eqnarray}
and mixed amplitudes
\begin{eqnarray}
a^{(0)}(p_i) &\leftrightarrow & \eta_i^0,\bar\eta_i^1~~~a^{(+\frac{1}{2})}(p_i)\leftrightarrow \eta_i^1,\bar\eta_i^0;\nonumber\\
a^{*(0)}(p_i) &\leftrightarrow & \eta_1^1,\bar\eta_i^0~~~a^{(-\frac{1}{2})}(p_i)\leftrightarrow \eta_i^0,\bar\eta_i^1.
\end{eqnarray}
Supersymmetry invariance requires that:
\begin{equation}
p^{\alpha \dot \beta}A_4=q^{\alpha}A_4=\bar q^{\dot \alpha}A_4=0.
\end{equation}
These constraints can be solved as
\begin{equation}
A_4(\Omega_1,\Omega_2,\Omega_3,\Omega_4)\sim \delta^4(p^{\alpha \dot \beta})\delta^2(q^{\alpha})\mathcal{C}_4,
\end{equation}
for the chiral amplitudes and as
\begin{equation}
A_4(\Omega_1,\Omega_2,\bar \Omega_3,\bar \Omega_4)\sim \delta^4(p^{\alpha \dot \beta})\delta^2(q^{\alpha})\delta^2(\bar q^{\dot\alpha}) \mathcal{M}_4,
\end{equation}
for the mixed amplitudes. Here $\mathcal{C}_4$ and $\mathcal{M}_4$ are purely bosonic functions that depend on external kinematics $\{\lambda_i,\tilde \lambda_i\}_{i=1}^n$ and the coupling constant. The fermionic delta functions are given by:
\begin{equation}
\delta^2(q^{\alpha})=\sum_{i,j=1}^n\langle ij \rangle ~\eta_i\eta_j,~\delta^2(\bar q^{\alpha})=\sum_{i,j=1}^n\langle ij \rangle ~\bar \eta_i\bar\eta_j.
\end{equation}
Comparing with the tree level expression for the component amplitude (here we consider all momenta as ingoing ones and  omit the spinor indices)
\begin{equation}
A_4^{tree}(\psi_1,\psi_2,\phi_3,\phi_4)\sim \delta^4\left(\sum_{i=1}^4\lambda_i\tilde\lambda_i\right)\langle 12\rangle,
\end{equation}
we conclude that $\mathcal{C}_4^{tree}=1$ and
\begin{equation}
A_4^{tree}(\Omega_1,\Omega_2,\Omega_3,\Omega_4)=g \delta^4\left(\sum_{i=1}^4\lambda_i\tilde\lambda_i\right)
\delta^2\left(\sum_{i=1}^4\lambda_i\eta_i\right).
\end{equation}
Thus, the polarisation factor for the chiral amplitude is simply $\delta^4(p^{\alpha \dot \beta})\delta^2(q^{\alpha})$ and this amplitude to all loop orders can be written as:
\begin{equation}
A_4^{tree}(\Omega_1,\Omega_2,\Omega_3,\Omega_4)=g \delta^4\left(\sum_{i=1}^4\lambda_i\tilde\lambda_i\right)
\delta^2\left(\sum_{i=1}^4\lambda_i\eta_i\right)~C(s,t,u,g).
\end{equation}

As for the mixed amplitude,  at the tree level it is equal to zero, i.e. $\mathcal{M}^{tree}_4=0$. Performing a similar comparison with the component one-loop computations, one can conclude that at the one loop level:
\begin{equation}
A_4^{1-loop}(\Omega_1,\Omega_2,\bar\Omega_3,\bar\Omega_4)=g^2 \delta^4\left(\sum_{i=1}^4\lambda_i\tilde\lambda_i\right)
\delta^2\left(\sum_{i=1}^4\lambda_i\eta_i\right)\delta^2\left(\sum_{i=1}^4\tilde\lambda_i\bar\eta_i\right) I^{(1)}_1,
\end{equation}
where $I^{(1)}_1$ is the $s$-channel one-loop bubble integral. The amplitudes with other positions of $\bar\Omega_i$ are given by the same expression but with the bubble integral in different kinematic channels. Thus, in this case, the polarisation factor from (\ref{schemAmpl}) is given by $\delta^4(p^{\alpha \dot \beta})\delta^2(q^{\alpha})\delta^2(\bar q^{\dot\alpha})$ and the corresponding amplitude can be written to all loop orders as:
\begin{equation}
A_4(\Omega_1,\Omega_2,\bar\Omega_3,\bar\Omega_4)=g^2 \delta^4\left(\sum_{i=1}^4\lambda_i\tilde\lambda_i\right)
\delta^2\left(\sum_{i=1}^4\lambda_i\eta_i\right)\delta^2\left(\sum_{i=1}^4\tilde\lambda_i\bar\eta_i\right) M(s,t,u,g).
\end{equation}

\newpage
\bibliographystyle{utphys2}
\bibliography{refs}

\end{document}